\RequirePackage{shellesc}

\documentclass[sigconf]{acmart}

\AtBeginDocument{%
  }

\setcopyright{acmlicensed}
\acmJournal{PACMMOD}
\acmYear{2025} \acmVolume{3} \acmNumber{1 (SIGMOD)} \acmArticle{65} \acmMonth{2}\acmDOI{10.1145/3709715}

\acmConference[Conference acronym 'XX]{Make sure to enter the correct
  conference title from your rights confirmation emai}{June 03--05,
  2018}{Woodstock, NY}
\acmISBN{978-1-4503-XXXX-X/18/06}

\usepackage{algorithm2e}
\usepackage{colortbl}
\usepackage{enumitem}
\usepackage{multirow}
\usepackage{tikz}
\usetikzlibrary{shapes, arrows.meta, positioning}

\newtheorem{problem}{Problem}

\begin{document}
\title{Progressive Entity Resolution: A Design Space Exploration}

\author{Jakub Maciejewski}
\orcid{0009-0005-8307-8843}
\affiliation{%
  \institution{National and Kapodistrian University of Athens}
    \country{Greece}
}
\email{sdi1700080@di.uoa.gr}

\author{Konstantinos Nikoletos}
\orcid{0000-0003-3465-1197}
\affiliation{%
  \institution{National and Kapodistrian University of Athens}
    \country{Greece}
}
\email{k.nikoletos@di.uoa.gr}

\author{George Papadakis}
\orcid{0000-0002-7298-9431}
\affiliation{%
  \institution{National and Kapodistrian University of Athens}
    \country{Greece}
}
\email{gpapadis@di.uoa.gr}

\author{Yannis Velegrakis}
\orcid{0000-0001-5109-3700}
\affiliation{ 
\institution{University of Trento \& Utrecht University} 
    \country{The Netherlands}
}
\email{i.velegrakis@uu.nl}

\begin{abstract}
Entity Resolution (ER) is typically implemented  as a batch task that processes all available data before identifying duplicate records. However, applications with time or computational constraints, e.g., those running in the cloud, require a progressive approach that produces results in a pay-as-you-go fashion. Numerous algorithms have been proposed for Progressive ER in the literature. In this work, we propose a novel framework for Progressive Entity Resolution that organizes relevant techniques into four consecutive steps: (i) filtering, which reduces the search space to the most likely candidate matches, (ii) weighting, which associates every pair of candidate matches with a similarity score,  (iii) scheduling, which prioritizes the execution of the candidate matches so that the real duplicates precede the non-matching pairs, and  (iv) matching, which applies a complex, matching function to the pairs in the order defined by the previous step. We associate each step with existing and novel techniques, illustrating that our framework overall generates a superset of the main existing works in the field. We select the most representative combinations resulting from our framework and fine-tune them over 10 established datasets for Record Linkage and 8 for Deduplication, with our results indicating that our taxonomy yields a wide range of high performing progressive techniques both in terms of effectiveness and time efficiency.
\end{abstract}

\maketitle

\section{Introduction}

Entity Resolution (ER), often also referred to as Record Linkage (a.k.a. Clean-clean ER) or Deduplication (a.k.a. Dirty ER), is a fundamental task in data management \cite{DBLP:conf/sigmod/MudgalLRDPKDAR18}. It deals with the challenge of identifying and linking data structures, typically referred to as \textit{entity profiles}, that represent the same real-world object \cite{DBLP:series/synthesis/2015Dong}. The linked structures are referred to as duplicates. Detecting the duplicates is crucial for boosting the performance of a wide range of data management tasks, from recommendation to question answering.

An ER application in order to be successful, needs to overcome a series of challenges \cite{DBLP:series/synthesis/2015Dong,DBLP:books/daglib/0030287,DBLP:journals/pvldb/GetoorM12}. 
The first is \textit{Volume}. ER algorithms should scale to thousands or even millions of entity profiles. This isn't straightforward, due to the the inherently quadratic nature of ER, which is why it is addressed through filtering~\cite{DBLP:journals/csur/PapadakisSTP20}. The second is \textit{Variety}. ER should seamlessly apply to data sources that vary in format, schema and structure. Yet, most ER approaches are crafted either for structured, e.g., relational data or semi-structured, e.g., RDF~\cite{DBLP:series/synthesis/2021Papadakis}. The third is \textit{Velocity}. Most ER solutions operate in an offline mode, i.e., they produce results only after having processed all the input data. Unfortunately, there are cases where there are restrictions and cost in computational or temporal resources. An example is the cloud environment. In these cases, partial results are required within a specific time frame~\cite{DBLP:journals/tkde/SimoniniPPB19,DBLP:conf/sigmod/GalhotraFSS21,DBLP:journals/vldb/GalhotraFSS21,DBLP:conf/edbt/GazzarriH23}.

To address these challenges, we focus on \textit{Schema-agnostic Progressive Entity Resolution}. 
Velocity is addressed by the progressive functionality, which yields results before processing all input data through a pay-as-you-go functionality. Volume is addressed by Filtering, which restricts the computational cost to the most similar entity profiles, disregarding those dissimilar. Variety is addressed by the schema-agnostic functionality, which represents every entity profile through a concatenation of all attribute values, regardless the respective attribute names.

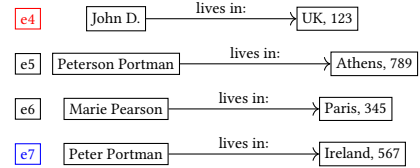
\begin{figure}[t]
{\scriptsize
\centering
\centering
\begin{tabular}{|c|c|c|c|c|}
\hline
\rowcolor{cyan!30}
id & First name & Last name & City & Zip Code \\
\hline
\textcolor{red}{e1} & John & Doe & London & 123 \\
\hline
\textcolor{blue}{e2} & Peter & Portman & Dublin & 567 \\
\hline
e3 & Donald & Hirsh & New Jersey & 486 \\
\hline
\multicolumn{5}{c}{\textbf{(a) Data source $D_1$}}
\end{tabular}
\centering
\vspace{12pt}
\begin{tikzpicture}[node distance=0.6cm, auto]
    \node[rectangle, draw=black, draw=red, align=center] (id4) {\textcolor{red}{e4}};
    \node[rectangle, draw=black, align=center, below of=id4] (id5) {e5};
    \node[rectangle, draw=black, align=center, below of=id5] (id6) {e6};
    \node[rectangle, draw=black, draw=blue, align=center, below of=id6] (id7) {\textcolor{blue}{e7}};
    \node[rectangle, draw=black, align=center, right=of id4] (e4) {John D.};
    \node[rectangle, draw=black, align=center, below of=e4] (e5) {Peterson Portman};
    \node[rectangle, draw=black, align=center, below of=e5] (e6) {Marie Pearson};
    \node[rectangle, draw=black, align=center, below of=e6] (e7) {Peter Portman};
    \node[rectangle, draw=black, align=center, right=2cm of e4, anchor=west] (e4_country) {UK, 123};
    \node[rectangle, draw=black, align=center, right=2cm of e5, anchor=west] (e5_country) {Athens, 789};
    \node[rectangle, draw=black, align=center, right=2cm of e6, anchor=west] (e6_country) {Paris, 345};
    \node[rectangle, draw=black, align=center, right=2cm of e7, anchor=west] (e7_country) {Ireland, 567};
    \draw[->] (e4) -- node[midway, above] {lives in:} (e4_country);
    \draw[->] (e5) -- node[midway, above] {lives in:} (e5_country);
    \draw[->] (e6) -- node[midway, above] {lives in:} (e6_country);
    \draw[->] (e7) -- node[midway, above] {lives in:} (e7_country);
\end{tikzpicture}\\
\vspace{-12pt}
\textbf{(b) Data source $D_2$}\\
\begin{center}
\begin{tabular}{c}
\begin{minipage}{0.5\textwidth}
\centering
\textcolor{red}{(e1, e4)},~\hspace{12pt}  \textcolor{blue}{(e2, e7)}, ~\hspace{12pt}~(e2, e5), ..., ~\hspace{12pt}(e3, e6)
\end{minipage} \\
\textbf{(c) Ordered set of candidate pairs}\\
\end{tabular}
\end{center}
\vspace{-10pt}
\caption{Two clean data sources and the candidate pairs generated by a Progressive Entity Resolution approach.}
\label{fig:example}
}
\end{figure}

Consider  Figure~\ref{fig:example} that illustrates a Record Linkage case, i.e., a case of two data sources $D_1$ and $D_2$, each one being duplicate-free, meaning  it contains no matching entity profiles, nevertheless, across the sources there are duplicates. For instance, $e_1$ matches to $e_4$ and $e_2$ to $e_7$. The two sources need to be merged, and to do so, the duplicates need to be detected. Furthermore, the two data sources exhibit high variety, varying both in  format (structured vs semi-structured) and in  schema. Typically, an ER process consists of three steps~\cite{DBLP:books/daglib/0030287,DBLP:journals/pvldb/GetoorM12}: Filtering, Entity Matching and Clustering. Yet, due to the small size of this example, Filtering is omitted, while Clustering lies out of the scope of our work. Therefore, we exclusively focus on Entity Matching, where a typical batch algorithm considers all pairs of entities, returning $e_1 \equiv e_4$ and $e_2 \equiv e_7$ after processing the entire data sources, i.e., after examining 12 pairs. In contrast, Progressive Entity Resolution defines a processing order, as in Figure \ref{fig:example}(c), so as to promote the most likely matches and detect most of them even if the processing is terminated prematurely.

Most existing progressive methods are independent of matching decisions, defining a \textit{static} processing order \cite{DBLP:journals/tkde/SimoniniPPB19,DBLP:journals/tkde/WhangMG13,DBLP:journals/tkde/PapenbrockHN15}. Yet, they have some limitations: 
\begin{enumerate}[leftmargin=*]
    \item They disregard recent advances in Filtering: they exclusively consider (meta-)blocking as a pre-processing step for addressing volume, but recent studies show that nearest neighbor search achieves a much higher performance \cite{DBLP:journals/pvldb/PaulsenGD23,DBLP:journals/pvldb/Thirumuruganathan21,neuhof2024open}.
    \item They disregard recent advances in natural language processing and the dominance of the semantic entity representations that stem from pre-trained language models in both ER tasks, i.e., blocking and matching. Instead, they exclusively rely on the syntactic representation of entity profiles \cite{DBLP:journals/pvldb/Thirumuruganathan21,DBLP:journals/pvldb/ZeakisPSK23,neuhof2024open,DBLP:conf/adma/SunJXSNW23,wang2024pre}.
    \item There is no generic framework that unifies the traditional, syntactic- and blocking-based approaches with the latest, semantic-based ones that leverage nearest neighbor search.
\end{enumerate}

To address these shortcomings, we propose a novel framework for Progressive Entity Resolution that is generic enough to cover any type of entity representation and of Filtering methods. Our goal is not to combine the outcome of different techniques (e.g., the syntactic- and the semantic-based), but to propose a set of generic, integrated steps that offer a series of options to researchers and practitioners, facilitating the construction of progressive pipelines that leverage diverse methods in a seamless way. To the best of our knowledge, no other framework is offering this in a unifying way.

Our design space consists of four consecutive steps: 
\begin{enumerate}[leftmargin=*]
    \item Filtering restricts the computational cost to the most similar pairs of candidates. 
    \item Weighting assigns a similar score to each candidate pair that is proportional to the matching likelihood of the entity profiles comprising it.
    \item Scheduling leverages the similarity scores in order to define the optimal processing order that gives precedence to the duplicate pairs over the non-matching ones.
    \item Matching analytically examines each candidate pair to decide whether its entity profiles are matching or not. 
\end{enumerate}

We explain how the filtering step of our framework incorporates all the existing state-of-the-art approaches, from blocking- and sorting-based  to those leveraging nearest neighbor search. For each approach, we discuss the corresponding weighting functions for syntactic and semantic representations. For scheduling, we propose four main algorithms based on the concept of the similarity graph \cite{DBLP:series/synthesis/2021Papadakis}, which includes a node for each input entity with an edge connecting each candidate pair. We stress that the existing Progressive Entity Resolution approaches cover only a small portion of those generated by our framework. We also stress that the matching process lies outside the scope of this work, since it is an orthogonal issue~\cite{DBLP:journals/tkde/SimoniniPPB19,DBLP:journals/tkde/WhangMG13}, with numerous recent state-of-the-art solutions based on Deep Learning~\cite{DBLP:journals/sigmod/FanTLWDJGT24,DBLP:conf/adma/SunJXSNW23}.

To test the performance of our framework, we perform a grid search to fine-tune a wide range of end-to-end pipelines over a set of 10 well-established real-world datasets for Record Linkage and of 8  for Deduplication. The experimental results indicate the best combination of filtering and scheduling algorithms in terms of effectiveness. We compare the best configuration of each filtering approach with the state-of-the-art from the literature, i.e., DeepBlocker \cite{DBLP:journals/pvldb/Thirumuruganathan21} and Sparkly \cite{DBLP:journals/pvldb/PaulsenGD23}. The former leverages semantic representations and the latter syntactic. This is the first time the two algorithms are applied on Progressive Entity Resolution. Our results indicate that DeepBlocker consistently underperforms our semantic-based nearest neighbor approach both with respect to effectiveness and to time efficiency. 
Sparkly is much faster than our syntactic-based nearest neighbor approach (due to its Spark-based parallelization), but at the same time, it is significantly less accurate. We also consider an additional state-of-the-art techniques as baseline method, namely
I-PES~\cite{DBLP:conf/edbt/GazzarriH23}.

Overall, we make the following contributions:

\begin{enumerate}[leftmargin=*]
    \item We introduce a generic framework for Progressive Entity Resolution consisting of four steps, that organize the existing approaches, and giving raise to some new. Among the latter are the first progressive methods leveraging nearest neighbor search.
    \item We analytically explain the configuration space of each progressive method generated by our framework. Some of the new methods are the first  to apply pre-trained language models to Progressive Entity Resolution. 
    \item We perform an extensive experimental analysis of all Progressive Entity Resolution approaches and configurations (using grid search) over 18 well-established datasets. Our experiments give valuable insights into the relative performance of the filtering and scheduling algorithms and demonstrate the superiority of our approaches over the state-of-the-art in the literature.
\end{enumerate}

The implementation of our approaches is available online at:  \url{https://github.com/JacobMaciejewski/PER-Design-Space-Exploration}.

\section{Related Work}
\label{sec:relatedWork}

Progressive Entity Resolution methods can be distinguished in two main categories, the \textit{static} and the \textit{dynamic}~\cite{DBLP:series/synthesis/2021Papadakis}.
The former  generate a processing order that is independent of the matching decisions, while the latter update the processing order based on the latest matching decision(s). Our work focuses on the former for three reasons. First, they offer a more realistic setting, where scheduling is performed once, without the need of an oracle for the matching. The dynamic assume an oracle with 100\% matching accuracy in every turn. Second, the static are more efficient than the dynamic, because the dynamic rearrange the candidate pairs after each matching decision. This raises the computational cost significantly. Last, but not least, most existing Progressive Entity Resolution methods yield a static processing order~\cite{DBLP:journals/tkde/WhangMG13,DBLP:journals/tkde/SimoniniPPB19}. 

Our design space organizes the static methods proposed in~\cite{DBLP:journals/tkde/WhangMG13,DBLP:journals/tkde/SimoniniPPB19} into a unified framework that facilitates their extension and comparison. It also integrates the latest works in blocking and nearest neighbor search~\cite{DBLP:conf/icde/0001FSMAN23,DBLP:journals/pvldb/ZeakisPSK23}, which correspond to the Filtering step and partially to the Weighting step of our design space. State-of-the-art approaches like Sparkly~\cite{DBLP:journals/pvldb/PaulsenGD23} and DeepBlocker~\cite{DBLP:journals/pvldb/Thirumuruganathan21} can also be integrated into our framework, however, they are used as baseline methods in our experimental analysis. 

Static approaches over dynamic data have recently been studied~\cite{DBLP:conf/edbt/GazzarriH23} by applying Progressive Entity Resolution to streaming data that is not available upfront, but arrives at varying rates. The goal is to identify duplicates soon after their arrival, while scaling to large volumes. The specific framework gives rise to three different schema-agnostic algorithms, of which the entity-centric I-PES consistently exhibits the highest performance, and is, consequently, experimentally compared to our approaches in Section~\ref{sec:sota}.

Unlike the static progressive, the dynamic progressive rely on a perfect matching algorithm in order to iteratively rearrange the processing order of the candidate pairs. The Dynamic Progressive Sorted Neighborhood~\cite{DBLP:journals/tkde/PapenbrockHN15} organizes the sorted entities into a two dimensional array such that after detecting a match in $A(i, j)$, the processing moves on to check $A(i+1, j)$ / and $A(i, j+1)$. This is a dynamic extension of our sorting-based workflows. The \textit{pBlocking}~\cite{DBLP:journals/vldb/GalhotraFSS21} is another method that initially generates a set of blocks, that is then iteratively refined through block cleaning based on the ratio of duplicate and non-duplicate profiles as it is determined after a limited amount of matching decisions in every round. 
Comparison cleaning based on meta-blocking is also applied. This approach, which has been demonstrated in the \textit{BEER system} \cite{DBLP:conf/sigmod/GalhotraFSS21}, is a dynamic extension of the blocking workflows of our design space. 

Disk-based dynamic methods~\cite{DBLP:journals/access/SunHSN22} have been used in cases of extremely large datasets that cannot be entirely loaded into memory. 
They aim to avoid high I/O overhead by optimally scheduling the transition of data between main memory and the hard disk, while searching for the most promising pairs and defining their processing order. They leverages a cost benefit analysis to split data into partitions that are iteratively scheduled for processing. In Query-driven ER~\cite{DBLP:journals/pvldb/SimoniniZBN22}, the ER is performed on query results rather than entire datasets. A data lake is queried and the produced results are progressively returned after deduplication. This idea has been implemented in the BrewER system~\cite{DBLP:journals/pvldb/ZecchiniSBN23}.

It should be stressed that the blocking workflows of our design space are based on the pipeline presented elsewhere~\cite{DBLP:journals/pvldb/0001SGP16}, but go beyond it by including Block Filtering~\cite{neuhof2024open} and many more weighting schemes (ref. to W1-W14 in Section~\ref{sec:blockingWkfl}). Furthermore, the specific work~\cite{DBLP:journals/pvldb/0001SGP16}  focuses exclusively on batch ER, missing the Scheduling phase which renders a blocking workflow suitable for Progressive Entity Resolution. To the best of our knowledge, no prior work examines the performance of progressive workflows that combine a diverse set of blocking and weighting techniques with novel scheduling algorithms through a thorough experimental analysis.

Due to the Scheduling, Active Learning based ER~\cite{DBLP:conf/kdd/SarawagiB02,DBLP:journals/pvldb/JainSS21,DBLP:journals/pacmmod/GenossarGS23}, AL-based for short, is similar to Progressive Entity Resolution because it also assigns scores to the candidate pairs generated by Filtering. However, instead of trying to place the matching pairs before the non-matching, AL-based ER aims to minimize the number of labeled pairs that are required for training a supervised classifier for Entity Matching. To this end, it operates iteratively, selecting in each iteration the most critical unlabeled pairs, whose labeling will reinforce the distinction between the positive and the negative classes (e.g., the most uncertain pairs as determined through a committee of lassifiers)~\cite{DBLP:series/synthesis/2021Papadakis,DBLP:conf/sigmod/Meduri0SS20}. In other words, AL-based ER promotes pairs with a matching likelihood close to 0.5, unlike Progressive Entity Resolution, which promotes pairs with a matching likelihood close to 1.0. Using the former in the place of the latter (or vice versa) would result in poor performance. In the future, it is worth exploring more elaborate techniques for combining AL-based ER with Progressive Entity Resolution.

\section{Problem Statement}

Entity Resolution  comes in different forms depending on the number of input data sources~\cite{DBLP:journals/pvldb/GetoorM12,DBLP:books/daglib/0030287,DBLP:series/synthesis/2015Dong,DBLP:journals/csimq/SaeediNPR18,DBLP:journals/csur/ChristophidesEP21}. The main  are: 
\begin{itemize}[leftmargin=*]
    \item \textit{Deduplication}, often referred to as \textit{Dirty ER}, receives as input a single data source $D$ that contains duplicates. The goal is to partition $D$ into sets of duplicate entity profiles such that every set corresponds to a different real-world object.
    \item \textit{Record Linkage}, often referred to as \textit{Clean-Clean ER}, receives as input two data sources $D_1$ and $D_2$, with each one containing no dublicates in itself. Its goal is to identify the duplicate profiles across the two sources. 
    \item \textit{Multi-source ER} generalizes Record Linkage to more than two duplicate-free, but overlapping data sources. The goal is to cluster together the duplicate profiles from the different sources. This task can also be treated as a series of Record Linkage tasks or as a single Deduplication task, where the profiles of the identified linkages are merged into one after each step. 
\end{itemize}

In all cases, the solution to ER typically consists of the two consecutive steps forming the \textit{Filtering-Verification} framework \cite{DBLP:journals/csur/PapadakisSTP20,DBLP:conf/sigmod/MudgalLRDPKDAR18,DBLP:journals/pvldb/Thirumuruganathan21}. First, Filtering reduces the search space to the most similar entity profiles, thus excluding the obvious non-matches. This is an approximate process that curtails the originally quadratic time complexity at the cost of missing a limited portion of the duplicate profiles. Its outcome comprises a set of candidate pairs, which is then processed by Verification: typically, a complex, time-consuming function is applied to each pair $<e_i, e_j>$, yielding either a binary decision (``match'' or ``non-match'') or a numeric score proportional to the matching likelihood of $e_i$ and $e_j$.

A major drawback of the Filtering-Verification framework is its \textit{batch} functionality, which yields results only after processing the entire input \cite{DBLP:series/synthesis/2021Papadakis}. This is not compatible with ER applications having strict performance requirements with respect to run-time and/or computational costs \cite{DBLP:journals/pvldb/PaulsenGD23,DBLP:journals/pvldb/Thirumuruganathan21,neuhof2024open}. For example, consider applying ER to large corporate data lakes that run on third-party cloud infrastructures, which charge according to the computational resources that are consumed (e.g., the AWS Lambda functions). In these settings, the preferred solution is a \textit{progressive} functionality that produces results in a pay-as-you-go manner \cite{DBLP:series/synthesis/2021Papadakis}. This requires that the matching pairs take precedence over the non-matching  such that the more processing time is available or the more verifications are performed, the more duplicates are detected. 

\begin{figure}[!t]
  \centering
  \includegraphics[width=0.50\linewidth]{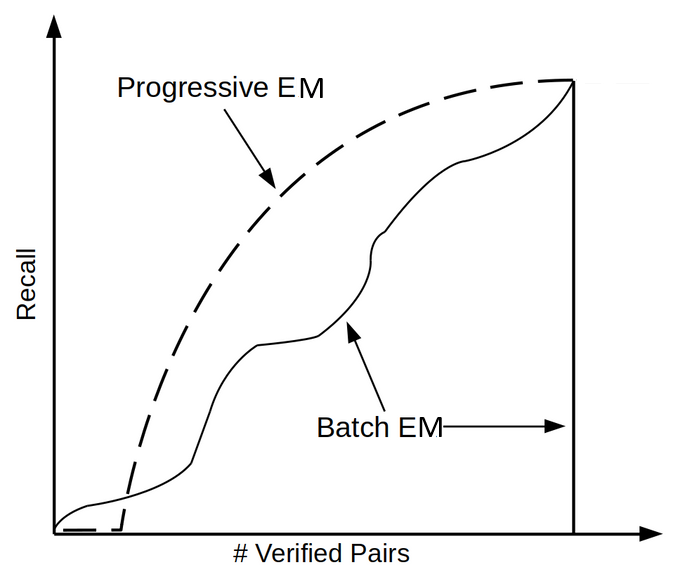}
  \vspace{-10pt}
  \caption{Batch vs Progressive Entity Resolution.}
  \label{fig:PAYGOER}
\end{figure}

More formally, assuming that a batch Entity Matching approach needs to verify $N$ candidate pairs in order to process a data source~$D$, a progressive one should satisfy the following requirements \cite{DBLP:journals/tkde/WhangMG13}: 
\begin{enumerate}[leftmargin=*]
    \item \textit{Higher early effectiveness}. If a batch and a Progressive Entity Resolution approach verify $N'$ candidate pairs such that $N'$$\ll$$N$, the latter detects many more matching pairs than the former.
    \item \textit{Same eventual effectiveness}. Upon verifying $N$ pairs, the Progressive Entity Resolution approach yields the same duplicates as the batch one. 
\end{enumerate}
The two requirements are  highlighted in Figure \ref{fig:PAYGOER}. The horizontal axis corresponds to the verified pairs, while the vertical one corresponds to recall. We define the area under the curve as \textbf{progressive recall$\mathbf{@N}$}, where $N$ is the budget of the maximum verified pairs. It takes values in $[0,1]$, with higher values indicating higher effectiveness. In other words, the higher the progressive recall is, the more and earlier are the existing duplicates detected.

In this context, our goal can be formally described as follows: 
\begin{problem}
Given two data sources, $D_1$ and $D_2$, along with a budget of $N$ verifications, \textbf{Progressive Entity Resolution} produces a set of candidate pairs ordered such that progressive recall$@N$ is maximized, while the run-time is minimized.
\end{problem}

Note that the above definition can be easily adapted to Deduplication and Multi-source ER as well as to a budget defined in terms of maximum run-time. Note also that the run-time excludes the Verification time \cite{DBLP:journals/tkde/SimoniniPPB19,DBLP:journals/tkde/WhangMG13} -- it only considers the time that intervenes between receiving the input data sources and returning the ordered set of candidate pairs. A similar assumption holds for progressive recall \cite{DBLP:journals/tkde/SimoniniPPB19,DBLP:journals/tkde/WhangMG13}: it is independent of Verification performance, assuming an oracle function that always decides with 100\% accuracy whether two entity profiles are matching or not.

\section{Design Space of Progressive Entity Resolution}

\begin{figure}[t]
  \centering
  \includegraphics[width=0.95\linewidth]{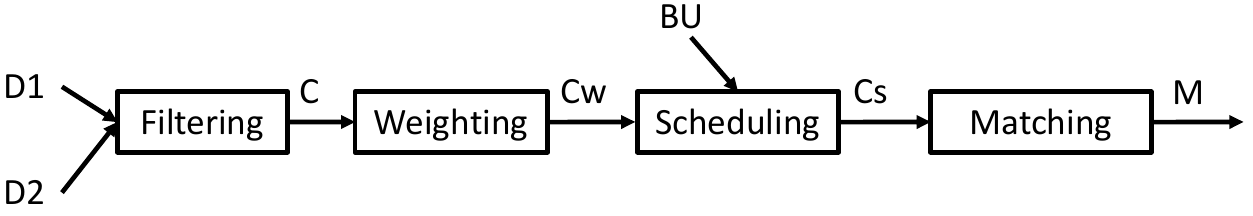}
  \vspace{-10pt}
  \caption{Progressive Entity Resolution workflow}
  \label{fig:workflows}
\end{figure}

We now propose an architecture template for Progressive Entity Resolution solution  consisting of four modules applied in the following order, also illustrated in Figure \ref{fig:workflows}:
\begin{enumerate}[leftmargin=*]
    \item \textit{Filtering} receives as input two clean data sources, $D_1$ and $D_2$, (or a single dirty one) and returns as output the set of candidate pairs $C$, which includes the most likely matches.
    \item \textit{Weighting} receives as input the set of candidates pairs $C$ defined by filtering. Its output comprises a set $C_w$ with the same pairs, where each one is assigned to a positive weight that is proportional to the matching likelihood of its entities.
    \item \textit{Scheduling} receives as input the user-defined budget of $BU$ verifications along with the weighted candidate pairs $C_w$ and defines their processing order in such a way that the most likely matches are examined first. Therefore, its output comprises $C_s$, which is a permutation of the given set $C_w$.
    \item \textit{Matching} receives $C_s$ as input and iteratively outputs to the next candidate pair to be verified. In other words, it simply applies a matching function to the pairs in $C_s$ according to the processing order defined by Scheduling. The set of verified pairs is denoted by $M$.
\end{enumerate}

For each module, we discuss a diverse approaches. First for the Filtering and Weighting (Section \ref{sec:filtering}), and then the Scheduling (Section~\ref{sec:scheduling}). An exception is the  Matching, which is common to both batch and progressive solutions, with a bulk of recent works combining language models with deep learning to achieve  high accuracy~\cite{DBLP:journals/pvldb/Thirumuruganathan21,DBLP:journals/sigmod/FanTLWDJGT24,DBLP:conf/icde/0001KCP24}. This is why Matching is out of the scope of this work. The same applies to the clustering step, which is necessary for an end-to-end ER pipeline. The integration of the state-of-the-art clustering techniques for Record Linkage~\cite{DBLP:journals/vldb/PapadakisETHC23} and Deduplication~\cite{DBLP:journals/pvldb/HassanzadehCML09,DBLP:journals/pvldb/VesdapuntBD14} in our framework is left for future work.

\section{Filtering \& Weighting}
\label{sec:filtering}

Due to the quadratic time complexity of ER, Filtering is necessary for curtailing the search space by discarding the apparent non-matches \cite{DBLP:journals/pvldb/PaulsenGD23,DBLP:books/daglib/0030287,DBLP:journals/pvldb/Thirumuruganathan21}. In other words, Filtering retains only the most likely matches, as determined by their high similarity, through a quick and approximate process of low time complexity, which can be accomplished in one of the following ways:

\begin{enumerate}[leftmargin=*]
\item \textbf{NN workflows.} The input entity profiles are embedded into dense, high-dimensional vectors through pre-trained language models. In Record Linkage, one of the two data sources is indexed, while the vectorized profiles of the other query the index to retrieve their semantically $k$ nearest neighbors. 
\item \textbf{Join workflows.} Unlike the semantic focus of NN workflows, the join workflows leverage syntactic similarities: they follow the same approach of indexing and querying, but convert the input entity profiles into sparse, multi-dimensional numeric vectors. Every dimension in these vectors corresponds to a different character or token $n$-gram, with a weight proportional to its frequency in the values of the respective profile.
\item \textbf{Blocking workflows}. Signatures, called blocking keys, are extracted from the attribute values of each profile. Each signature $s$ creates a separate block $b_s$ containing all entities associated with $s$. The resulting blocks are refined based on the assumption that the larger a block is, the less likely it is to contain distinctive information. The refined blocks are then converted into a graph, where block sharing is translated into node adjacency. The edges are weighted by metrics quantifying the block co-occurrence of the corresponding entity profiles. 
\item \textbf{Sorting-based Workflows.} They rely on the same signatures as blocking workflows, but define as candidates the pairs with similar (not identical) blocking keys. The entity profiles are alphabetically ordered, according to the signatures extracted from their attribute values. This results in a sorted list of entities, which is iteratively processed using sliding windows of increasing size. The matching likelihood of two profiles is proportional to their co-occurrence frequency in these windows.
\end{enumerate}

Following \cite{DBLP:journals/tkde/SimoniniPPB19}, all approaches operate in a schema-agnostic manner that disregards all attribute names, but considers all attribute values. None of them involves a learning-based functionality that requires a labelled dataset. Instead, they all rely on heuristics, trading high time efficiency at the cost of lower effectiveness. Below, we elaborate on their functionality along with the corresponding weighting functions and the configuration parameters.



\subsection{NN Workflows}
\label{sec:NNworkflows}

\begin{algorithm}[t]
{\small
\LinesNumbered
\caption{Outline of the NN and Join workflows.}\label{alg:NN-Workflow1}
    \KwIn{$\mathcal{D}_1$, $\mathcal{D}_2$:The data sources to be matched, $k$: the maximum number of candidates per query entity, $LM$: the Language Model that vectorizes each entity, $Sim$: the similarity function between two embedding vectors. $IS$: the indexing scheme. }
    \KwOut{$\mathcal{C}$: the resulting set of candidate pairs}
    \ForEach{entity $e_i \in \mathcal{D}_1$}{
        $\mathcal{E}(e_i) \gets \text{LM.Embed}(e_i)$\;
        $I$.index($\mathcal{E}(e_i)$)\;
    }

    $C \gets \{\}$ \tcp*{Set of candidate pairs}
    \ForEach{entity $e_j \in \mathcal{D}_2$}{
        $\mathcal{E}(e_j) \gets \text{LM.Embed}(e_j)$\;
        $\mathcal{C}_j \gets$ $I$.getNN($\text{Sim}$, $\mathcal{E}(e_j)$, $k$)\;
        $\mathcal{C} \gets \mathcal{C} \cup \mathcal{C}_i$\;
    }
    \KwRet{$\mathcal{C}$}\;
}
\end{algorithm}

The functionality of NN workflows is outlined in Algorithm \ref{alg:NN-Workflow1}. The input comprises the two data sources to be matched along with the maximum number of candidates per query entity, corresponding to its most probable matches. For $k$, we consider three representative values (i.e., $k \in \{1, 5, 10\}$). The input should also specify the pre-trained language model $LM$ that converts the given entities into embedding vectors. Following \cite{DBLP:journals/pvldb/ZeakisPSK23}, we consider 10 established LMs (we disregard AlBERT \cite{DBLP:conf/iclr/LanCGGSS20} and XLNet \cite{DBLP:conf/nips/YangDYCSL19}, due to their consistently high run-time and low effectiveness): 
\begin{itemize}[leftmargin=*]
    \item the main static ones, which always associate every token or character n-gram with the same precalculated embedding vector:\\ 
    Word2Vec \cite{DBLP:journals/corr/abs-1301-3781, DBLP:conf/nips/MikolovSCCD13}, FastText \cite{DBLP:journals/corr/BojanowskiGJM16} and  Glove \cite{DBLP:conf/emnlp/PenningtonSM14}.
    \item the main BERT-based ones, which offer a per-token contextual vectorization: BERT \cite{DBLP:conf/naacl/DevlinCLT19}, DistilBERT \cite{DBLP:journals/corr/abs-1910-01108}, and RoBERTa \cite{DBLP:journals/corr/abs-1907-11692}. 
    \item the main SentenceBERT-based ones, which associate the entire schema-agnostic representation of every entity with a single context-aware embedding vector: S-MiniLM \cite{DBLP:conf/nips/WangW0B0020}, S-MPNet \cite{DBLP:conf/nips/Song0QLL20}, S-GTR-T5 \cite{DBLP:journals/jmlr/RaffelSRLNMZLL20} and S-DistilRoBERTa \cite{DBLP:journals/corr/abs-1907-11692}.
\end{itemize}

Another configuration parameter is the similarity function $Sim: \mathbb{R} \rightarrow [0,1]$ between two multidimensional embedding vectors $v, w \\ \in \mathbb{R}^n$. We consider two established options:

\begin{enumerate}[leftmargin=*]
    \item The Euclidean one, which considers the magnitude and direction of the two vectors, but is sensitive to their dimensionality~\cite{DBLP:journals/pvldb/ZeakisPSK23}: $$s_{\text{euclidean}}(\mathbf{v}, \mathbf{w}) = \frac{1.0}{1.0 + d_{\text{euclidean}}(\mathbf{v}, \mathbf{w})} = \frac{1.0}{1.0 + \sqrt{\sum_{i=1}^{n}(v_i - w_i)^2}}$$
    \item The cosine similarity, which exclusively considers the angle between the two input vectors: $$s_{\text{cosine}}(\mathbf{v}, \mathbf{w}) = 1.0 - d_{\text{cosine}}(\mathbf{v}, \mathbf{w}) = \frac{\sum_{i=1}^{n}v_i \cdot w_i}{\sqrt{\sum_{i=1}^{n}v_i^2} \cdot \sqrt{\sum_{i=1}^{n}w_i^2}}$$
\end{enumerate} 

The final configuration parameter is the \textit{indexing scheme}, which exclusively applies to Record Linkage, designating which of the input data sources will be indexed -- leaving the other one as the query set. Three are the possible options: 
\begin{enumerate}[leftmargin=*]
    \item indexing the smallest source,
    \item indexing the largest one, or
    \item both.
\end{enumerate}

In Algorithm \ref{alg:NN-Workflow1}, we index the first data source $D_1$ and query with the second one, $D_2$. More specifically, every entity in $D_1$ is converted to the embedding vector of the given $LM$, after concatenating all its attribute values in a sentence without any special tokens \cite{DBLP:journals/pvldb/ZeakisPSK23} (Line 2). The resulting vector is indexed by a state-of-the-art tool for nearest neighbor search (Line 3). Based on \cite{DBLP:journals/is/AumullerBF20}, we use FAISS \cite{DBLP:journals/tbd/JohnsonDJ21} for this purpose, as it constitutes one of the fastest and most effective libraries for indexing high dimensional embedding vectors and retrieving the nearest neighbors per query. 

Next, all entity profiles in $D_2$ are vectorized one by one, using the same LM (Lines 6-7). Every embedding vector is posed as a query to the index $I$, returning the $k$ most similar entities from $D_1$ (Line 8). These candidates are added to the set of candidate pairs $C$, which is returned as output (Lines 9-11). Note that internally, $\mathcal{C}$ associates every candidate pair with its similarity score as determined by the given similarity function, $Sim$.

The time complexity of Algorithm \ref{alg:NN-Workflow1} depends on the time complexity of each query to the index $I$, i.e., $O(|D_2| \cdot |q_I(D_1)|)$, where $|q_I(D_1)|$ is the time complexity of a single entity on $I$, after having indexed $D_1$ entities. Note that $|q_I(D_1)|$ is constant and $|q_I(D_1)|\ll |D_1|$, due to FAISS' internal functionality, which partitions the indexed vectors in such a way that every query is restricted to a few partitions (rather than the entire indexed data source). Theoretically, the time complexity of vectorizing all input entity profiles is linear and, thus, lower than the cost of querying. Similarly, the space complexity of Algorithm \ref{alg:NN-Workflow1} is determined by the the cost of storing the vectors of $D_1$ entities in memory, i.e., $O(|D_1|)$.

NN workflows have not been applied to Progressive ER before.

\subsection{Join Workflows}

Approaches of this type follow Algorithm \ref{alg:NN-Workflow1}, involving two phases: 
\begin{enumerate}[leftmargin=*]
    \item the \textit{Indexing phase} in Lines 1-4, where one of the input datasets is transformed to a structure suitable for the fast detection of nearest neighbors, and
    \item the \textit{Querying phase} in Lines 5-10, where entities from the other input dataset query the indexing structure to detect their nearest neighbors.
\end{enumerate}
The only difference between Join and NN workflows lies in the vectorization approach: unlike the dense embedding vectors of the latter, which map entities to the semantic space of language models, the Join workflows leverage sparse multi-dimensional vectors directly extracted from the attribute values of the input entities. More specifically, two functions are combined to this end:

\begin{enumerate}[leftmargin=*]
    \item The \textit{tokenization function} converts the concatenated attribute values of each entity into a set of character or token n-grams: $n\in \{3, 4, 5\}$ in the former case and $n\in \{1, 2\}$ in the latter one.
    \item The \textit{feature scoring function} associates every dimension in the sparse vector with a numerical score. We consider 3 options: 
    \begin{enumerate}[leftmargin=*]
        \item Boolean scores (\textsf{BS}) indicate the presence or absence of an n-gram in the attribute values of the given entity. 
        \item Term-frequency scores (\textsf{TF}) indicate how many times each n-gram appears in the attribute values of the given entity. Note that the frequency of each n-gram is normalized by the highest frequency within the given entity.
        \item TF-IDF scores (\textsf{TF-IDF}) extends \textsf{TF} to encompass the n-gram's importance across the entire input dataset through the logarithmically scaled inverse fraction of the number of profiles that contain the n-gram.
    \end{enumerate}
\end{enumerate}
In the following, we consider all combinations of tokenization and feature scoring functions. Note that instead of FAISS, we use a standard inverted index for quickly retrieving the candidates per query entity, due to the sparse vectors used by Join workflows.
Note that Join workflows have not been applied to Progressive Entity Resolution before.

\subsection{Blocking Workflows}
\label{sec:blockingWkfl}

{\small
\begin{algorithm}[t]
\LinesNumbered
\caption{Outline of the blocking workflows.}\label{workflow:block-1}
    \KwIn{$\mathcal{D}_1$, $\mathcal{D}_2$:The data sources to be matched, $WS$: the weighting scheme}
    \KwOut{$\mathcal{G} = \mathcal{V} \times \mathcal{E}$: the similarity graph }
    $\mathcal{B} \leftarrow tokenBlocking(\mathcal{D}_1, \mathcal{D}_2)$\;
    $\mathcal{B}' \leftarrow blockPurging(\mathcal{B})$\;
    $\mathcal{B}'' \leftarrow blockFiltering(\mathcal{B}')$\;
    $\mathcal{G} \leftarrow \{\}$\;
    \ForEach{entity $e_i \in D$}{
        $B_{e_i} \gets \bigcup\limits_{b \in \mathcal{B}''}^{} b \mid e_i \in b$\;
        \ForEach{block $b \in B_{e_i}, e_i \in \mathcal{D}_1$}{
            \ForEach{$e_j \in b : e_j \in \mathcal{D}_2$}{
                $\mathcal{V} \gets \mathcal{V} \cup \{n_j\}$\;
                $\mathcal{E}  \gets \mathcal{E} \cup \{(n_i, n_j)\}$\;
                $weight(n_i, n_j)$ = $WS(n_i, n_j, \mathcal{B}'')$
            }
        }
    }
    \KwRet{$\mathcal{G}$}\;
\end{algorithm}
}

The functionality of these solutions is outlined in Algorithm \ref{workflow:block-1}.

Token Blocking \cite{DBLP:journals/pvldb/0001SGP16,neuhof2024open} is first applied (Line 1), generating a separate block for each token in the attribute values of the given entity profiles. Any other blocking method can be used, too, but Token Blocking is the only parameter-free one. 

Next, Block Purging \cite{neuhof2024open,neuhof2024open} is applied (Line 2) to remove oversized blocks, which comprise a large number of pairs, but very few (if any) are matching and have no other block in common. Similar to Token Blocking, this is a parameter-free approach. The core assumption is that \textit{the larger a block is, the more likely it is to contain repeated pairs that are not duplicates}. 

The same assumption lies at the core of the subsequent step (Line 3), Block Filtering \cite{neuhof2024open}: it removes every entity from a specific portion of its largest blocks, thus reducing the unnecessary pairs that involve non-matching entities at a small cost in recall.  Following \cite{DBLP:journals/is/PapadakisMGSTGB20}, this ratio is set to 80\%.

Based on the blocks resulting from the initial steps, a similarity graph $\mathcal{G}$ is created (Lines 4-14). This is an undirected graph, whose nodes correspond to entities and its edges connect the candidate pairs (Lines 9-10). Every edge is weighted according to the characteristics of blocks containing every one of the adjacent entities as well as the characteristics of common blocks (Line 11). These characteristics include the number of blocks, their \textit{size}, i.e., their total number of entities, and their \textit{cardinality}, i.e., their candidate pairs.

More specifically, the weighting scheme is the sole configuration parameter of the blocking workflows. Its rationale is similar to that of Block Purging and Block Filtering: the more and smaller blocks two entities share (i.e., the more and less frequent their common signatures are), the more likely they are to be matching. In this context, the possible weighting schemes for two entities, $e_i$ and $e_j$, are the following \cite{DBLP:journals/pvldb/0001SGP16,neuhof2024open}: 
\begin{enumerate}[label=W\arabic*), leftmargin=*]
    \item \textsf{Common Blocks}: $\mathbf{CB}=|B_i \cap B_j|$, where $B_x$ stands for the set of blocks containing entity $e_x$ and $|B_x|$ for its size.
    \item \textsf{Cosine}=$|B_i \cap B_j|/\sqrt{|B_i| \cdot |B_j|}$=$CB/\sqrt{|B_i| \cdot |B_j|}$.
    \item \textsf{Dice}=$2 \cdot |B_i \cap B_j|/(|B_i| + |B_j|)$=$2 \cdot CB /(|B_i| + |B_j|)$.
    \item \textsf{Jaccard}=$|B_i \cap B_j|/|B_i \cup B_j|$=$CB/(|B_i| + |B_j| - CB)$.    
    \item \textsf{Size Normalized Common Blocks}:\textbf{\textsf{SN-CB}}=$\sum_{b\in B_i \cap B_j}1/|b|$.
    \item \textsf{Size Normalized Cosine}: $SN$-$CB/\sqrt{SN\text{-}B_i \cdot SN\text{-}B_j}$, where $SN\text{-}B_i$ $= \sum_{b\in B} 1/|b|$, with SN denoting size normalization and $|b|$ symbolizing the number of entities in block $b$.
    \item \textsf{Size Normalized Dice}: $2\cdot SN$-$CB/(SN$-$B_i + SN$-$B_j)$.
    \item \textsf{Size Normalized Jaccard}: $SN$-$CB/(SN$-$B_i + SN$-$B_j - SN$-$CB)$.
    \item \textsf{Cardinality Normalized Common Blocks}: \textbf{\textsf{CN-CB}}=$\sum_{b\in B_i \cap B_j}1/||b||$.
    \item \textsf{Cardinality Normalized Cosine}: \textsf{CN-Cosine}=\\
    $CN$-$CB/\sqrt{CN\text{-}B_i \cdot CN\text{-}B_j}$, where $CN\text{-}B_i$ $= \sum_{b\in B} 1/||b||$, with CN denoting cardinality normalization and $||b||$ symbolizing the number of pairs in block $b$.
    \item \textsf{Cardinality Normalized Dice}: \textsf{CN-Dice}=$2\cdot CN$-$CB/(CN$-$B_i + CN$-$B_j)$.
    \item \textsf{Cardinality Normalized Jaccard}: \textsf{CN-Jaccard}=$CN$-$CB/(CN$-$B_i + CN$-$B_j - CN$-$CB)$.
    \item \textsf{Enhanced Common Blocks}: \textsf{ECB}=$CB \cdot \log\frac{|B|}{|B_j|} \cdot \log\frac{|B|}{|B_j|}$.
    \item \textsf{Enahnced Jaccard}: \textsf{EJS}=\textsf{Jaccard} $\cdot \log\frac{|V|}{|v_j|} \cdot \log\frac{|V|}{|v_i|}$.
\end{enumerate}

Note that the time and space complexity of Algorithm \ref{workflow:block-1} is determined by the number of pairs in the final set of blocks $\mathcal{B}''$, i.e., $O(||\mathcal{B}''||)$. Note also that only \textsf{CN-CB} has been applied to Progressive Entity Resolution before (it is called ARCS in \cite{DBLP:journals/tkde/SimoniniPPB19}).

\subsection{Sorting-based Workflows}

{\small
\begin{algorithm}[t]
\LinesNumbered
\caption{Outline of the sorting-based workflows.}\label{workflow:sn-1}
\KwIn{$\mathcal{D}_1$, $\mathcal{D}_2$:The data sources to be matched, $w$: the window size, 
    $WS$: the weighting scheme}
    \KwOut{$\mathcal{C}$: the resulting set of candidate pairs}
    $\mathcal{P} \gets \text{SortedNeighborhood}(\mathcal{D}_1, \mathcal{D}_2)$\;
    $C \gets \{\}$ \tcp*{Set of candidate pairs}
        \ForEach{entity $e_i \in \mathcal{D}_1$}{
            $N_i \gets \{\}$ \tcp*{Set of neighbors}
            $\text{positions}_i \gets \mathcal{P}\text{.getPositions}(e_i)$\;
            \ForEach{position $ps \in \text{positions}_i$}{
                $N_i \gets N_i \cup (\mathcal{P}\text{.getNeighbors}(ps, w) \cap \mathcal{D}_2)$\;
            }
            \ForEach{entity $e_j \in N_i$}{
                        $sim_{i,j} = \mathcal{P}\text{.getSimilarity}(e_i, e_j, WS) $\;
                        $\mathcal{C} \gets \mathcal{C} \cup (e_i, e_j, sim_{i,j})$\;
                        
            }
        }
    \KwRet{$\mathcal{C}$}\;
\end{algorithm}
}

The functionality of these solutions is outlined in Algorithm \ref{workflow:sn-1}. 

Initially, Sorted Neighborhood is applied (Line 1): first, it alphabetically sorts all tokens appearing in the attribute values of all input entities and then, it sorts in \textit{random} order the entities corresponding to each token \cite{DBLP:journals/tkde/SimoniniPPB19}. The resulting sorted list of entities is stored in an array $P$. A window $w$ slides over this list to detect the candidate pairs. Its size $w$ is fixed, given as a configuration parameter that should be at least 2 so that at least two entities co-occur in each window. We consider all integers in $[2,10]$.

Subsequently, for each input entity, we retrieve its positions in the array $P$ (Lines 3-5); each position $ps$ yields a neighborhood, which includes all other entities in positions $P[ps+1]$, $P[ps+2]$,..., $P[ps+w]$, where $w$ is the current size of the window. These entities are the candidate pairs of the current entity $e_i$. They are all placed in a set $N_i$ comprising the neighbors of $e_i$ (Lines 6-8). Note that the same entity might appear multiple times in the neighborhood of $e_i$, due to different, contiguous tokens. Every neighboring entity $e_j$ is then considered as a matching candidate of $e_i$ (Lines 9-12) based on a similarity that stems from the closeness of their associated positions. The weighted pairs are aggregated in the set of candidate pairs that is returned as output (Lines 11-14).

We consider the following options for the \textit{weighting scheme} $WS$ that is given as configuration parameter to compute the similarity between two candidate matches: 

\begin{enumerate}[leftmargin=*]
    \item Absolute Co-occurrence Frequency counts the number of positions that co-occur in the window of size $w$: $\mathbf{ACF}(e_i, e_j, w) = | \{ | ps_i - ps_j | < w : ps_i \in positions_i \wedge ps_j \in positions_j \} | $, where $positions_x$ is the set of positions associated with entity $e_x$.
    \item Normalized Co-occurrence Frequency, which is inspired from the Jaccard similarity:\\
    $NCF(e_i, e_j, w) = \frac{ACF(e_i, e_j, w)}{|positions_i| + |positions_j| - ACF(e_i, e_j, w)}$.
    \item Dice Normalized Co-occurrence Frequency:\\ $DNCF(e_i, e_j, w) = 2 \times \frac{ACF(e_i, e_j, w)}{|positions_i| + |positions_j|}$.
    \item Cosine Normalized Co-occurrence Frequency:\\ $CNCF(e_i, e_j, w) = \frac{ACF(e_i, e_j, w)}{\sqrt{|positions_i| \times |positions_j|}}$.
    \item Inverse Distance, which sums the inverse distances between two positions that are located within the same window $w$:\\ $ID(e_i, e_j, w)$ =  $\sum \frac{1}{| p_i - p_j |} \forall p_i \in positions_i,  p_j \in positions_j,\\ | p_i - p_j | < w $.
\end{enumerate}

Note that only NCF has already been applied to Progressive Entity Resolution (it is called RCF in \cite{DBLP:journals/tkde/SimoniniPPB19}). Note also that Algorithm \ref{workflow:sn-1} entails another parameter, the \textit{functionality scope}, which can be:
\begin{enumerate}[leftmargin=*]
    \item \textit{Local Scope} is the one presented in Algorithm \ref{workflow:sn-1}, emitting all candidate pairs for a particular, predetermined window size $w$. 
    \item \textit{Global Scope} repeats the processing in Lines 3-13 for a range of window sizes: from 2 to the given size $w$, which in this case sets the maximum value. In each iteration, the similarity scores of the candidate pairs identified in smaller windows are updated by adding the scores from the current window. 
\end{enumerate} 

The time complexity of Algorithm \ref{workflow:sn-1} is dominated by the computation of candidate pair weights and the sorting of the tokens appearing in the attribute values of the input entities, i.e., $O(|C| + |T|\cdot \log|T|)$, where $|T|$ is the number of these tokens (we expect $|D_1| \ll |T|$ and $|D_2| \ll |T|$, because each entity contains multiple tokens in its attribute values). The space complexity is dominated by the size of the array $P$ storing the sorted list of entities, $O(|P|)$.

\subsection{Application to Deduplication}

The above are crafted for Record Linkage, where the input comprises of two datasets $\mathcal{D}_1$ and $\mathcal{D}_2$, but can be easily adapted for the case of Deduplication, where the input is a single dataset $\mathcal{D}$ with duplicates in itself. For that, some minor changes are required in Algorithm \ref{alg:NN-Workflow1}. Line 1 needs to index $\mathcal{D}$, Line 6 to use all entities in $\mathcal{D}$ as queries and Line 9 to ensure that no duplicate pairs are added in the output set $\mathcal{C}$ (i.e., that pair $<$$e_i, e_j$$>$, $i\neq j$, does not appear in the form $<$$e_j, e_i$$>$). This can be accomplished by predetermining the place of every entity in every new pair $<$$e_x, e_y$$>$ added to $\mathcal{C}$, by requiring the entity with the lower id is placed on the left side of the pair. 
This unified form allows for automatic discardment of duplicate pairs, given that the $\mathcal{C}$ is a set. Algorithm \ref{workflow:block-1} also requires minor changes to adapt to Deduplication. Line 1 provides as input to Token Blocking only the $\mathcal{D}$, while the entities $e_i$ in Line 7 and $e_j$ in Line 8 should be different (i.e., $i \neq j$), as both belong to $\mathcal{D}$. Finally, Algorithm \ref{workflow:sn-1} requires that in Line 1, the SortedNeighborhood receives as input only $\mathcal{D}$, while Line 3 goes through all entities in $\mathcal{D}$. In Line 7, instead of ensuring that every neighbor $e_j$ is from a different data source, we need to ensure that it is different from $e_i$ (i.e., $j \neq i$). Finally, every pair $<$$e_i, e_j$$>$ added to the output set $\mathcal{C}$ in Line 11 should be ordered, similar to the Algorithm \ref{alg:NN-Workflow1} adaptation, i.e., $<$$e_x, e_y$$>$~$\Rightarrow$~$x$$<$$y$.

\section{Scheduling}
\label{sec:scheduling}

\begin{table*}[t]\centering
\footnotesize
\setlength{\tabcolsep}{2pt}
	\begin{tabular}{ | l | r | r | r | r | r | r | r | r | r | r || r | r | r | r | r | r | r | r | }
		\cline{2-19}
		\multicolumn{1}{c|}{}&
		\multicolumn{1}{c|}{D$_1$ {\scriptsize\cite{DBLP:conf/semweb/2010om}}} &
		\multicolumn{1}{c|}{D$_2$ {\scriptsize\cite{DBLP:journals/pvldb/KopckeTR10}}} &
		\multicolumn{1}{c|}{D$_3$ {\scriptsize\cite{DBLP:journals/pvldb/KopckeTR10}}} &
		\multicolumn{1}{c|}{D$_4$ {\scriptsize\cite{DBLP:journals/pvldb/KopckeTR10}}} &
		\multicolumn{1}{c|}{D$_5$ {\scriptsize\cite{DBLP:conf/esws/ObraczkaSR21}}} &
        \multicolumn{1}{c|}{D$_6$ {\scriptsize\cite{DBLP:conf/esws/ObraczkaSR21}}} &
        \multicolumn{1}{c|}{D$_7$ {\scriptsize\cite{DBLP:conf/esws/ObraczkaSR21}}} &
        \multicolumn{1}{c|}{D$_8$ {\scriptsize\cite{DBLP:conf/sigmod/MudgalLRDPKDAR18}}} &
        \multicolumn{1}{c|}{D$_9$ {\scriptsize\cite{DBLP:journals/pvldb/KopckeTR10}}} &
        \multicolumn{1}{c||}{D$_{10}$ {\scriptsize\cite{DBLP:journals/is/PapadakisMGSTGB20}}} & 
        \multicolumn{1}{c|}{De$_1$ {\scriptsize\cite{DBLP:journals/pvldb/VesdapuntBD14}}} & 
        \multicolumn{1}{c|}{De$_2$ {\scriptsize\cite{DBLP:journals/tkde/PapenbrockHN15}}} & 
        \multicolumn{1}{c|}{De$_3$ {\scriptsize\cite{DBLP:journals/pvldb/VesdapuntBD14}}} & 
        \multicolumn{1}{c|}{De$_4$ {\scriptsize\cite{DBLP:journals/pvldb/0001SGP16}}} & 
        \multicolumn{1}{c|}{De$_5$ {\scriptsize\cite{DBLP:journals/pvldb/0001SGP16}}} & 
        \multicolumn{1}{c|}{De$_6$ {\scriptsize\cite{DBLP:journals/pvldb/0001SGP16}}} & 
        \multicolumn{1}{c|}{De$_7$ {\scriptsize\cite{DBLP:journals/pvldb/0001SGP16}}} &
        \multicolumn{1}{c|}{De$_8$ {\scriptsize\cite{DBLP:journals/pvldb/0001SGP16}}} \\
		\hline
        \hline
        D$_{\alpha}$ & Rest.1 & Abt  & Amazon & DBLP & IMDb & IMDb & TMDb & Walmart &  DBLP & IMDb & \multirow{ 2}{*}{Cora} & \multirow{2}{*}{CDdb} & \multirow{ 2}{*}{Product} & \multirow{ 2}{*}{10K} & \multirow{ 2}{*}{50K} & \multirow{ 2}{*}{100K} & \multirow{ 2}{*}{200K} & \multirow{ 2}{*}{300K}\\
        D$_{\beta}$ & Rest.2 & Buy & GB & ACM & TMDb & TVDB & TVDB & Amazon & GS & DBpedia & & & & & & & & \\
        \hline
        $|$D$_{\alpha}$$|$ 
        & 339 & 1,076 & 1,354 & 2,616 & 5,118 & 5,118 & 6,056 & 2,554 & 2,516 & 27,615 & \multirow{ 2}{*}{1,878} & \multirow{ 2}{*}{2,161} & \multirow{ 2}{*}{9,763} & \multirow{ 2}{*}{10$^4$} & \multirow{ 2}{*}{5$\cdot$10$^4$} & \multirow{ 2}{*}{10$^5$} & \multirow{ 2}{*}{2$\cdot$10$^5$} & \multirow{ 2}{*}{3$\cdot$10$^5$}\\
        $|$D$_{\beta}$$|$ 
        & 2,256 & 1,076 & 3,039 & 2,294 & 6,056 & 7,810 & 7,810 & 22,074 & 61,353 & 23,182 & & & & & & & & \\
        \hline
		$|Dup|$ & 89 & 1,076 & 1,104 & 2,224 & 1,968 & 1,072 & 1,095 & 853 & 2,308 & 22,863 & 62,892 & 1,085 & 299 & 8,705 & 43,071 & 85,497 &  172,403 & 257,034 \\
     CP & 7.7$\cdot10^5$ & 1.2$\cdot10^6$ & 4.1$\cdot10^6$ &  6.0$\cdot10^6$ & 3.1$\cdot10^7$ & 4.0$\cdot10^7$ & 4.7$\cdot10^7$ & 5.6$\cdot10^7$ &  1.5$\cdot10^8$ & 6.4$\cdot10^8$ & 1.8$\cdot10^6$ & 2.3$\cdot10^6$ & 4.8$\cdot10^7$ & 5.0$\cdot10^7$ & 1.3$\cdot10^9$ & 5.0$\cdot10^9$ & 2.0$\cdot10^{10}$ &  4.5$\cdot10^{10}$ \\
		\hline
	\end{tabular}
    \caption{Overview of the Record Linkage and Deduplication data sets used in our experiments, with $|$D$_x|$ denoting the number of entities in data source D$_x$, $|Dup|$ the number of duplicates in the groundtruth and CP the Cartesian product.
    }
\label{tb:ccerDatasets}
\end{table*}

\begin{table}[t]
    \centering
    \footnotesize
    \begin{tabular}{|l|l|}
        \hline
        \multicolumn{1}{|c|}{\textbf{Filtering type}} & 
        \multicolumn{1}{c|}{\textbf{Parameter Values}} \\
        \hline
        \hline
         \multirow{2}{*}{Common parameters} & scheduling algorithm $\in$ \{DFS, BFS, TOP, HB\} \\
            &  budget $\in \{i \times |Dup|, i \in [1,10] \}$\\
        \hline
        \hline
       \multirow{4}{*}{NN workflows} & indexing scheme $\in$ \{smallest, largest, both\} \\
        & similarity function $\in$ \{cosine, Euclidean\}\\
        & number of nearest neighbors $\in [1,5,10]$\\
        & language model $\in$ \{The 10 models in Sec. \ref{sec:NNworkflows}\} \\
        \hline
        \hline
        \multirow{7}{*}{Join workflows} &  indexing scheme $\in$ \{smallest, largest, both\} \\
        & similarity function $\in$ \{cosine, Euclidean\} \\
        &  number of nearest neighbors $\in [1,5,10]$ \\
        & weighting scheme $\in$ \{BW, TF, TF-IDF\} \\
        &  tokenizer $\in$ \{word unigrams, word bigrams, \\
        & character n-grams $n \in [3,5]$ \\
        \hline
        \hline
        Blocking workflows &  weighting scheme $\in$ \{{All WS from section \ref{sec:blockingWkfl}}\} \\
        \hline
        \hline
        Sorting- & window size $\in [1,2, \ldots, 10]$ \\
        based &  weighting scheme $\in$ \{ACF, NCF, DNCF, CNCF, ID\} \\
        workflows & functionality scope $\in$ \{local, global\} \\
        \hline
    \end{tabular}
    \caption{Configuration parameters per module.}
    \label{tb:configurations}
\end{table}

The goal of Scheduling is to define the optimal processing order of the weighted candidate pairs produced by the two previous modules in Figure \ref{fig:workflows}. Ideally, all matching pairs precede all non-matching ones. To this end, Scheduling receives as input the budget $BU$ specified by the user along with the set of weighted candidate pairs $C_w$ from Algorithms \ref{alg:NN-Workflow1} and \ref{workflow:sn-1} or the similarity graph from Algorithm \ref{workflow:block-1}. The two forms of input are equivalent and actually the former is transformed into the latter by creating an undirected graph $G=(V, E)$, where there is a separate node in $V$ for every input entities, while the edges in $E$ connect the candidate pairs and are weighted according to the respective similarity score.

Based on the similarity graph, the scheduling algorithms are distinguished into those focusing on edges or nodes. The former operate at a global level, considering all pairs, and the latter, at a local one, operating at the level of neighborhoods (i.e., they define a separate processing order for the candidates of each entity). More specifically, we introduce the following scheduling algorithms:

\begin{enumerate}[leftmargin=*]
    \item \textbf{Edge-centric Scheduling} (\textsf{EC}). It defines a global processing order by sorting all pairs in decreasing weight so as to retain the $BU$ top ranked ones. Its time complexity is $O(|E| \log |E|)$, but its space complexity is restricted to $O(|V| + |E|)$, because it suffices to use a priority queue that always contains the top-$BU$ weighted pairs ($BU \ll |E|$).
    \item \textbf{Node-centric Scheduling.} First, it assigns a score to each node, which is equal to the average similarity in its neighborhood. Then, it sorts the nodes in decreasing score. Finally, it orders the edges of each node neighborhood in decreasing weight. Its overall time complexity is $O(|V| (\log |V| + |\bar{N}| \log |\bar{N}|))$, where $\bar{N}$ is the average size of a node neighborhood, while its space complexity is $O(|V| + |E|)$.
    There are two variants in defining the final processing order of the candidate pairs:
    \begin{itemize}[leftmargin=*]
        \item \textbf{Depth-First Search} (\textsf{DFS}) starts with the top-weighted node, prioritizing all its edges in decreasing weight, then does the same with the next weighted node and so on. 
        \item \textbf{Breadth-First Search} (\textsf{BFS}) iteratively goes through the sorted list of nodes and in each round, it prioritizes the next top-weighted edge of the current node, if any.
    \end{itemize}
    \item \textbf{Hybrid.} It combines the operation of all the above algorithms. First, it computes the average node neighborhood per entity. During this process, it keeps in memory the best edge/candidate pair per neighborhood. These edges are globally sorted in decreasing weight. This is equivalent to applying \textsf{BFS} for one iteration. If the budget is not exhausted after processing these top-weighted pairs, it applies \textsf{DFS}: it sorts all nodes in decreasing average weight and, starting with the top-weighted one, it prioritizes all edges in its neighborhood, except for the top-weighted one, which has already been processed. Then, it moves to the next weighted node and so on. The time and space complexities are the same as in node-centric scheduling. 
\end{enumerate}

Note that the similarity graph is bipartite in the case of Record Linkage. This means that for the node-centric and the hybrid algorithms, it suffices to weight and sort only the nodes of one partition (this also ensures that no repeated pairs are included in the output of Scheduling). Note also that among these algorithms, only the Hybrid one has already been applied to Progressive Entity Resolution (see Progressive Profile Scheduling~in~\cite{DBLP:journals/tkde/SimoniniPPB19}).

\section{Experimental Analysis}

The goal of our experimental evaluation is threefold:
\begin{enumerate}[leftmargin=*]
    \item  to identify most effective solutions per filtering type.
    We discuss the performance of NN, join, blocking and sorting-based workflows in Sections \ref{sec:expNN}, \ref{sec:expJoins}, \ref{sec:expMB} and \ref{sec:expSN}, respectively;
`\item to assess the relative performance of the single best solution per filtering type in terms of progressive recall, run-time and memory consumption. This is examined in Section \ref{sec:overallBest};
    \item to compare the overall best solution of our architecture template with four state-of-the-art baseline methods, namely DeepBlocker \cite{DBLP:journals/pvldb/Thirumuruganathan21}, Sparkly \cite{DBLP:journals/pvldb/PaulsenGD23}, Progressive Block Scheduling \cite{DBLP:journals/tkde/SimoniniPPB19} and I-PES \cite{DBLP:conf/edbt/GazzarriH23}, with respect to progressive recall, run-time and memory footprint both in Record Linkage and in Deduplication. 
    This analysis is performed in Section~\ref{sec:sota}.
\end{enumerate}

\begin{table*}[t]\centering
{\scriptsize
\setlength{\tabcolsep}{4pt}
	\begin{tabular}{ | l | l | l | l | c || l | l | l | l | c || c ||  l | c | l |}
		\hline
        \multicolumn{1}{|c|}{Scheduling} &
        \multicolumn{1}{c|}{Indexing} &
		\multicolumn{1}{c|}{Similarity} &
        \multicolumn{1}{c|}{Language} &
        \multicolumn{1}{c||}{\multirow{2}{*}{$k$}} &
		\multicolumn{1}{c|}{Similarity} &
        \multicolumn{1}{c|}{\multirow{2}{*}{Tokenizer}} &
        \multicolumn{1}{c|}{Weighting} &
        \multicolumn{1}{c|}{Indexing} &
        \multicolumn{1}{c||}{\multirow{2}{*}{$k$}} &
        \multicolumn{1}{c||}{Weighting} & 
        \multicolumn{1}{c|}{Weighting} &
        \multicolumn{1}{|c|}{Window} &
        \multicolumn{1}{|c|}{Functionality} \\
		\multicolumn{1}{|c|}{Algorithm} &
        \multicolumn{1}{c|}{Scheme} &
		\multicolumn{1}{c|}{Function} &
        \multicolumn{1}{c|}{Model} &
        & \multicolumn{1}{c|}{Function} &
        &
        \multicolumn{1}{c|}{Scheme} &
        \multicolumn{1}{c|}{Scheme} &
        & \multicolumn{1}{c||}{Scheme} & 
        \multicolumn{1}{c|}{Scheme} &
        \multicolumn{1}{|c|}{Size} &
        \multicolumn{1}{|c|}{Scope} 
        \\
		\hline
        \hline
        EC & largest & Euclidean & S-GTR-T5 & 5 & Cosine & token unigram & TF-IDF & largest & 10 & \textbf{CN-CBS} & \textbf{ID} & \textbf{10} & \textbf{Global}\\
        \textbf{BFS} & \textbf{both} & \textbf{Euclidean} & \textbf{S-GTR-T5} & \textbf{5} & \textbf{Euclidean} & \textbf{character 5-gram} & \textbf{TF-IDF} & \textbf{both} & \textbf{5} & CBS & ACF & 10 & Global \\
        DFS & both & Cosine & S-GTR-T5 & 1 & Cosine & token unigram & TF-IDF & both & 1 & SN-CBS & DICE & 1 & Global \\
        Hybrid & both & Euclidean & S-GTR-T5 & 5 & Euclidean & character 5-gram & TF-IDF & both & 5 & CBS & ACF & 10 & Global\\
		\hline
        \multicolumn{1}{c}{} & \multicolumn{4}{c}{(a) NN workflows} &  \multicolumn{5}{c}{(b) Join workflows} & \multicolumn{1}{c}{(c) blocking workflows} & \multicolumn{3}{c}{(d) sorting-based workflows} \\
	\end{tabular}
    \caption{The best progressive solutions generated by our design space. The best per Filtering type are highlighted in bold.}
\label{tb:nnConf}
}
\end{table*}

\begin{figure*}[t]
  \centering
  \includegraphics[width=0.96\linewidth]{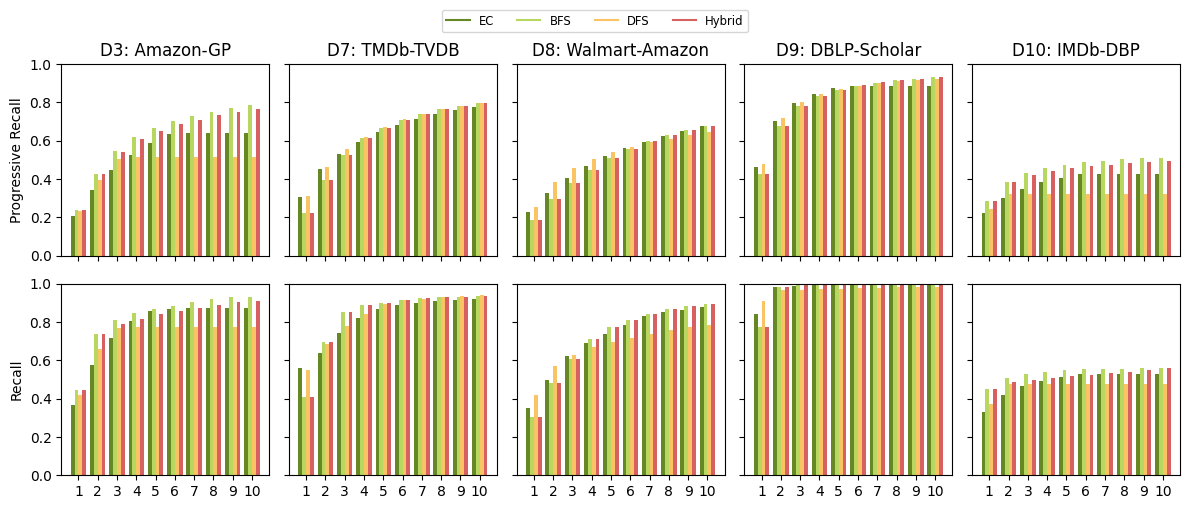}
  \vspace{-10pt}
  \caption{Progressive recall and recall of the best NN workflows in Table \ref{tb:nnConf}(a) across all budgets over selected datasets.}
  \label{fig:nnAUC}
\end{figure*}

\subsection{Experimental Setup}

All experiments were implemented in Python version 3.8.  
All experiments were executed on a server running Ubuntu 22.04 LTS, equipped with 64GB RAM, an Intel Core i7-9700K @ 3.60GHz and an NVIDIA GeForce RTX 2080.
The technical characteristics of the datasets used in our experiments are reported in Table \ref{tb:ccerDatasets} in increasing order of computational cost in terms of the Cartesian product (i.e., last line of the table). They include 10 real datasets for Record Linkage and 8 real and synthetic ones for Deduplication, all of which are popular in the literature \cite{DBLP:journals/pvldb/Thirumuruganathan21,DBLP:conf/sigmod/MudgalLRDPKDAR18,DBLP:journals/pvldb/KopckeTR10,DBLP:journals/pvldb/0001SGP16}, and described in the Appendix. 

In each dataset, we consider 10 different budgets, $BU_1, \ldots, BU_{10}$, where $BU_n = n \times |Dup|$, with $|Dup|$ denoting the number of duplicates in the corresponding dataset. For each budget, we perform grid search, considering all solutions generated by architectural template. Table~\ref{tb:configurations} summarizes the considered solutions per filtering type. There are 180 and 270 different solutions of NN and join workflows, respectively, while the blocking and sorting-based workflows yield 14 and 100, respectively. These solutions are combined with the four different scheduling algorithms presented in Section \ref{sec:scheduling}. Due to lack of space, we cannot report the performance of all solutions generated in this way. Instead, for each filtering type, Sections \ref{sec:expNN}-\ref{sec:expSN} report only the best solution per Scheduling algorithm with respect to the \textit{average distance from the top}.

More specifically, for each dataset, budget and filtering type, we first estimate the maximum progressive recall across all considered solutions and then, we estimate the distance of each solution from this maximum recall. We call this measure ``\textit{distance from the top}'' and formally define it as: $DFT = 1 - PR(so)/PR_{max}$, where $PR(so)$ is the progressive recall of solution $so$ and $PR_{max}$ the overall maximum value. Then, we estimate the average DFT of each solution across all datasets and budgets. Sections \ref{sec:expNN}-\ref{sec:expSN} discuss the performance of the solution per scheduler with the lowest average $DFT$.

\begin{figure*}[t]
  \centering
  \includegraphics[width=0.96\linewidth]{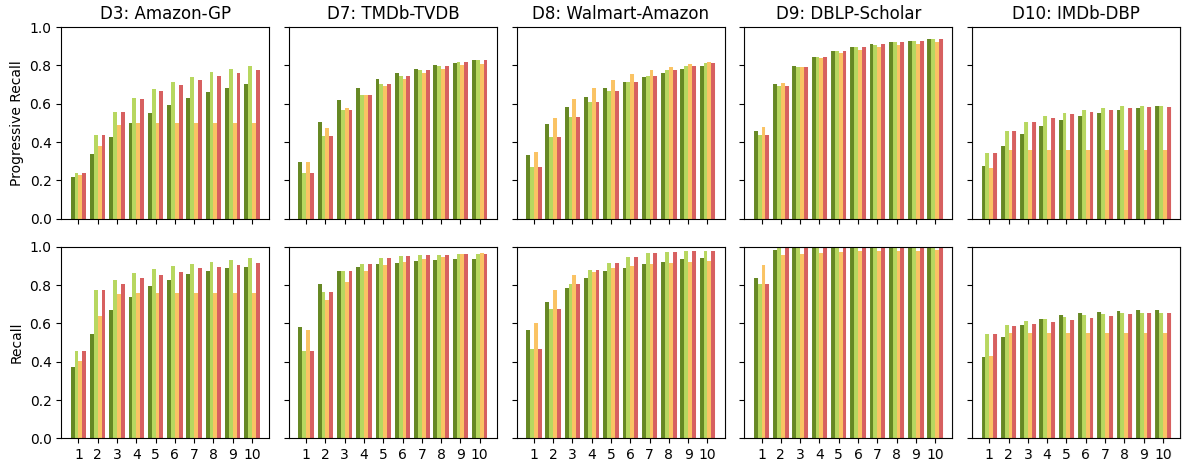}
  \vspace{-10pt}
  \caption{Progressive recall and recall of the best join workflows in Table \ref{tb:nnConf}(b) across all budgets over selected datasets.}
  \label{fig:joinAUC}
\end{figure*}

\subsection{NN workflows}
\label{sec:expNN}

The best NN workflows per scheduling algorithm across all datasets and budgets are reported in Table \ref{tb:nnConf}(a).
They all employ the S-GTR-T5 language model, which is identified as the most effective one in \cite{DBLP:journals/pvldb/ZeakisPSK23}, too. Most solutions combine the Euclidean similarity with 5 candidates per query entity. The only exception is \textsf{DFS}, where Cosine similarity takes a minor lead as long as a single candidate is returned per query entity. This essentially means that Scheduling is applied to the nearest neighbor per input entity, thus rendering the depth search inapplicable. Finally, most solutions index and query both data sources, a configuration that is more robust in all datasets. Only \textsf{EC} exclusively indexes the largest data source, using the smallest one as a query set. For this scheduling algorithm, the end result is practically identical with that of indexing both data sources, due to the large number of repeated candidate pairs generated by the latter, which are eliminated when \textsf{EC} merges them in a globally sorted candidate set.

The performance of these solutions with respect to progressive recall and to recall across all budgets in $D_3$ and $D_7$-$D_{10}$ is reported in Figure \ref{fig:nnAUC}. The performance for the rest of the datasets is presented in Figure \ref{fig:nnAUC2} in the Appendix, together with the detailed memory requirements and the run-times in Figures \ref{fig:nnMemory} and \ref{fig:nn_time}, respectively. The memory footprint is around 1 GB in all cases, as it is dominated by the high dimensional embedding vectors of the S-GTR-T5 model, while the differences in the run-time are insignificant, as it remains below 10 seconds in almost all cases.

Progressive recall yields two different patterns. In most datasets, $D_2$-$D_6$ and $D_{10}$, the \textsf{BFS} solution is the top performer, with the \textsf{Hybrid} one following in close distance: in half the cases, their average DFT, across all budgets, is practically identical, while their difference in $D_2$, $D_3$ and $D_{10}$ is less than 3\%. The \textsf{EC} solution consistently ranks third in these datasets except $D_4$, with an average progressive recall lower than \textsf{BFS} by 15\% to 21\%. The \textsf{DFS} underperforms all other algorithms, with its $DFT$ increasing with the increase of the budget. This should be expected, because it identifies a single candidate per input entity, failing to provide more candidates, despite the largest budget.

This situation is reversed the remaining four datasets, i.e., $D_1$ and $D_7$-$D_9$. In these datasets, the number of duplicates is much lower than the total number of input entities. As a result, the candidates gathered by \textsf{DFS} suffice for maximizing the progressive recall. \textsf{EC} follows in close distance, with \textsf{BFS} and \textsf{Hybrid} exhibiting practically identical performance, ranking last.

\begin{figure*}[h]
  \centering
  \includegraphics[width=0.96\linewidth]{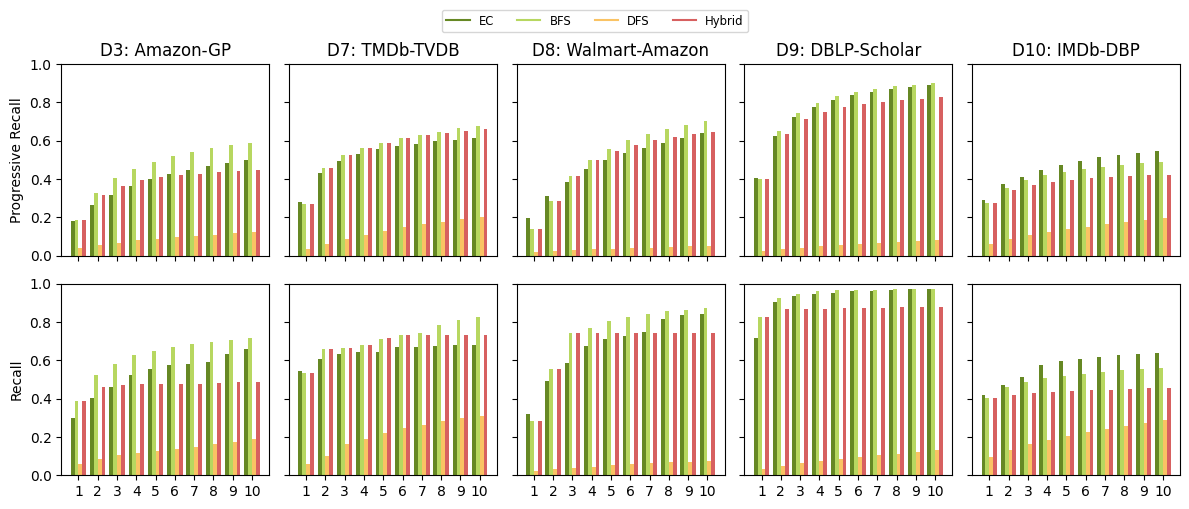}
  \vspace{-10pt}
  \caption{Progressive recall and recall of the best blocking workflows in Table \ref{tb:nnConf}(c) across all budgets over selected datasets.}
  \label{fig:blockAUC}
\end{figure*}

The above patterns apply to all budgets in the corresponding dataset, i.e.,
the relative performance of the considered solutions is consistent across the 10 budgets in each dataset. This means that in general, there is a high correlation between \textsf{DFS} and \textsf{EC} and a stronger one between \textsf{BFS} and \textsf{Hybrid}. The overall best NN solution applies the \textsf{BFS} algorithm to the 5 most similar candidates for each input entity according to the Euclidean similarity and the S-GTR-T5 embedding vectors. To this attests the relative performance of the NN solutions with respect to recall: \textsf{BFS} is consistently the top performer in half the datasets ($D_2$, $D_3$, $D_5$, $D_6$, $D_{10}$) typically followed by \textsf{Hybrid}, \textsf{EC} and \textsf{DFS} (in that order). Only in $D_1$, the situation is reversed, with \textsf{DFS} taking the lead. Yet, the differences are much smaller than those in terms of progressive recall, being insignificant in $D_4$, $D_7$, $D_8$, $D_9$.

\subsection{Join workflows}
\label{sec:expJoins}

The best join workflows per scheduling algorithm across all datasets and budgets are reported in Table \ref{tb:nnConf}(b). They all use TF-IDF, as it conveys more information than BW and TF. The \textsf{BFS} and the \textsf{Hybrid} solutions have the same configuration, whereas the \textsf{EC} and the \textsf{DFS} differ only in the indexing scheme and the number of nearest neighbors. The latter indexes and queries with both data sources to ensure robustness, while the former indexes only the largest data source, because its global sorting yields the same results as indexing both sources. \textsf{DFS} considers only the nearest neighbor of per profile, performing no depth search in practice. 

The progressive recall of these join solutions per scheduling algorithm across all budgets in $D_3$ and $D_7$-$D_{10}$ as well as the corresponding recall are reported in Figure \ref{fig:joinAUC} (ref to Figure~\ref{fig:joinAUC2} in the Appendix for the rest of the datasets alongside the memory requirements and the run-times in Figures~\ref{fig:joinMemory} and~\ref{fig:join_time}, respectively. Typically, the memory footprint does not exceed 100 MB, with \textsf{EC} being the only approach that does not index both data sources, thus requiring significantly lower memory. The run-times exhibit minor differences, remaining far below 10 seconds.

Regarding progressive recall, in five datasets ($D_2$, $D_3$, $D_5$, $D6$ and $D_{10}$), the \textsf{BFS} solution exhibits the highest values across all budgets, followed in close distance ($\ll2\%$) by the \textsf{Hybrid} configuration. The \textsf{EC} configuration consistently ranks third, with an average $DFT$ that ranges from 8\% ($D_{10}$) to 16\% ($D_3$). The \textsf{DFS} configuration consistently underperforms all others, typically falling short of the maximum progressive recall by at least~20\%.

The remaining datasets verify the very high correlation between \textsf{BFS} and \textsf{Hybrid}, especially during the lower budgets, where they basically yield the same candidate pairs, due to the identical configuration. Both rank second in $D_4$ and $D_7$, where \textsf{EC} is the top performer, with \textsf{DFS} ranking last, lower by 1/3, on average. The situation is reversed in $D_1$ and $D_8$, where the \textsf{DFS} configuration outperforms all others, leaving \textsf{BFS} and \textsf{Hybrid} in the last place.

Regarding recall, the differences between the four join solutions are consistently much lower than that of progressive recall. In four datasets ($D_4$ and $D_7$-$D_9$), there is actually an insignificant difference between them across all budgets. In the other datasets, \textsf{BFS} takes a clear lead, with \textsf{Hybrid} typically following in close distance, while \textsf{EC} and \textsf{DFS} are usually ranked third and fourth, resp. The only exception is $D_1$, where \textsf{DFS} and \textsf{EC} are the top performers.

Overall, the best join solution applies the \textsf{BFS} algorithm to the five most similar candidates per entity according to the Euclidean similarity between the character 5-grams vectors with TF-IDF weights.

\subsection{Blocking workflows}
\label{sec:expMB}

The best blocking workflows per scheduling algorithm across all datasets and budgets are reported in Table \ref{tb:nnConf}(c). Note that all weighting schemes rely on the number of blocks shared by two entities. In half the cases, normalization (by size or cardinality) is also required to increase the distinctiveness and the accuracy of the weights.

$D_3$ and $D_7$-$D_{10}$ is reported in Figure \ref{fig:blockAUC}. The performance over the other datasets appears in Figure~\ref{fig:blockAUC2} in the Appendix, both the memory requirements and the running time, in Figures~\ref{fig:pesmMemory} and~\ref{fig:pesm_time}, respectively. There are insignificant differences among the four solutions with respect to the memory footprint, given that they basically differ in the weighting scheme, while their run-time mostly depends on the dataset at hand.

We observe that the \textsf{DFS} configuration consistently underperforms all others to a significant extent. It achieves its best performance in $D_1$ and $D_5$, where its progressive recall is lower than the maximum one by 15\% and 22\%, on average, across all budgets, respectively. In all other datasets, its average $DFT$ exceeds 70\% -- in $D_8$ and $D_9$, this distance raises to a whole order of magnitude. Similar patterns pertain to recall, too.

The \textsf{Hybrid} configuration ranks third in most datasets with respect to both evaluation measures. In $D_1$ and $D_5$, it exhibits the worst performance among all schedulers, while in $D_7$ and $D_8$ it follows the top performer (\textsf{BFS}) in close distance.

There is a strong competition between the remaining solutions, the \textsf{EC} and the \textsf{BFS} one, both of which excel in 5 datasets. The former scores the highest progressive recall across all budgets in $D_1$, $D_2$, $D_5$, $D_6$ and $D_{10}$, and the latter in the rest. However, the difference between \textsf{EC} and \textsf{BFS} is much higher in the datasets, where the former ranks first: on average, \textsf{EC} underperforms \textsf{BFS} by less than 2\% in $D_4$ and $D_9$ as well as by less than 9\% in $D_7$ and $D_8$, whereas \textsf{BFS} underperforms \textsf{EC} by more than 10\% in $D_2$, $D_5$ and~$D_6$. Note that these patterns apply both to recall and progressive recall.

Overall, \textsf{EC} in combination with the \textsf{CN-CBS} weighting scheme constitutes the overall best blocking solution.

\begin{figure*}[t]
  \centering
  \includegraphics[width=0.96\linewidth]{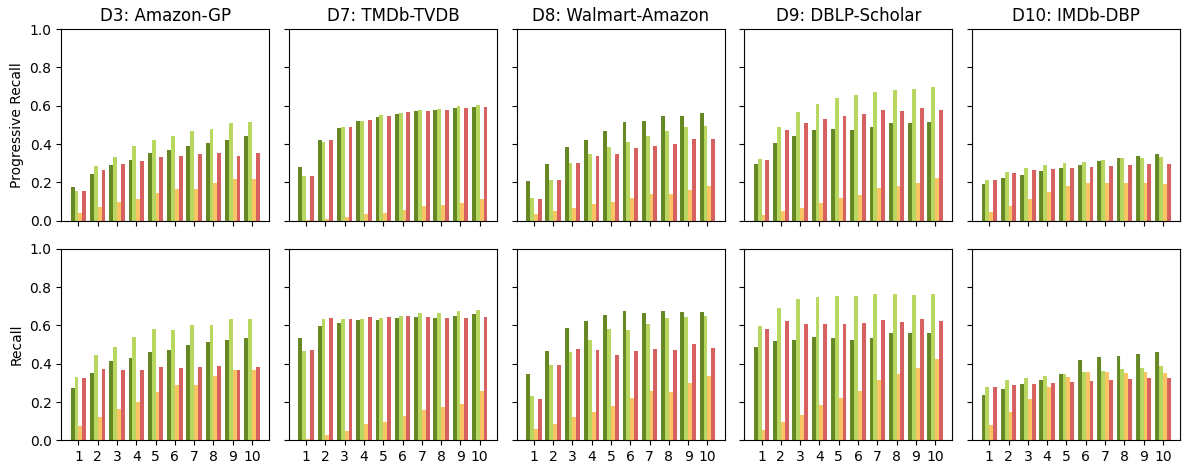}
  \vspace{-10pt}
  \caption{Progressive recall and recall of the best sorting-based workflows in Table \ref{tb:nnConf}(d) across all budgets over selected datasets.}
  \label{fig:snAUC}
\end{figure*}

\subsection{Sorting-based workflows}
\label{sec:expSN}

The best sorting-based solutions per scheduling algorithm across all datasets and budgets are reported in Table \ref{tb:nnConf}(d). They are all combined with the global scope of functionality. This indicates that considering the combined evidence from multiple windows performs better than the relying on a single window. Note that the largest window size works best for most scheduling algorithms, allowing for a larger difference between the values of two matching entities. 

The effectiveness of these solutions across all budgets over $D_3$ and $D_7$-$D_{10}$ is reported in Figure \ref{fig:snAUC}. The effectiveness over the other datasets is reported in Figure~\ref{fig:snAUC2} in the Appendix, while the memory consumption and the run-time per budget and dataset is reported in Figures \ref{fig:snMemory} and \ref{fig:sn_time}, respectively. In both cases, the differences are insignificant, due to the consistently low run-time (<10 sec) and memory footprint (<100 MB) -- only \textsf{DFS} is slightly more memory efficient, due to smallest window size.

Regarding progressive recall, the \textsf{DFS} configuration consistently underperforms all others to a significant extent, which raises to a whole order of magnitude in $D_7$. In the best case, in $D_5$, its progressive recall is lower than the highest one by 1/3, on average, across all budgets. Similar patterns apply to recall: \textsf{DFS} ranks last in 7 datasets, with its average $DFT$ ranging from 15\% ($D_2$) to~83\%~($D_7$).

On the other extreme lie the \textsf{EC} and the \textsf{BFS} solutions, with the former being the top performer in six datasets and the latter in the rest -- this applies to both evaluation measures. In terms of progressive recall, \textsf{EC} takes a major lead over \textsf{BFS} in half the datasets, with a progressive recall higher by at least 10\%. The only exception is $D_7$, where their difference is slightly above 1\%. In terms of recall, the difference between the two solutions is consistently smaller, but raises above 30\% in favor of \textsf{EC} in $D_1$, $D_5$ and $D_6$.

Finally, the behavior of \textsf{Hybrid} depends on the evaluation measure. For progressive recall, it consistently ranks third in all datasets where \textsf{EC} outperforms \textsf{BFS} (i.e., $D_1$-$D_2$ and $D_5$-$D_8$) -- yet, its progressive recall is much higher than that of \textsf{DFS}. In all other datasets, it ranks second, as its performance is highly correlated with \textsf{BFS}. For recall, it performs well only in datasets where all solutions have similar performance, namely $D_4$ andn $D_7$. In all other cases, its average $DFT$ ranges from 13\% ($D_6$) to 41\% ($D_1$).

Note that above patterns apply to each dataset, regardless of the budget. In other words, the relative performance of the four scheduling algorithms is not altered as the size of the budget increases. We can conclude, therefore, that \textsf{EC} in combination with the \textsf{ID} weighting scheme, a window size of 10 and the global functionality scope constitutes the overall best sorting-based solution.

\subsection{Overall best solution}
\label{sec:overallBest}

The above experiments demonstrate that for the NN and  join workflows, the \textsf{BFS} normally works best in datasets with high levels of noise (i.e., large portion of missing values) and high portion of duplicates for at least one the data sources. This applies to $D_2$, $D_3$, $D_5$, $D_6$ and $D_{10}$. In datasets where the matching entities share quite distinctive information, all Scheduling algorithms exhibit high performance, with minor differences between them. This applies to the bibliographic datasets, $D_4$ and $D_{10}$, due to the long, distinctive attribute values characterizing each entity (e.g., titles and author lists), as well as to $D_7$, due to the common movie titles and director or actor names (despite the high levels of noise). In datasets with low portion of duplicates for both data sources, i.e., $D_1$ and $D_8$, \textsf{DFS} is the top performer, when configured to search for the nearest neighbor per entity after indexing both data sources.

For the blocking and sorting-based workflows, the \textsf{BFS} and \textsf{EC} consistently outperform the other Scheduling algorithms. The former performs slightly better in datasets with duplicates sharing  highly distinctive information (i.e., $D_4$, $D_7$ and $D_9$), whereas \textsf{EC} works best in all other datasets, which exhibit with high levels of noise and/or low portion of duplicates for at least one data source.

\begin{figure*}[h]
  \centering
  \includegraphics[width=0.96\linewidth]{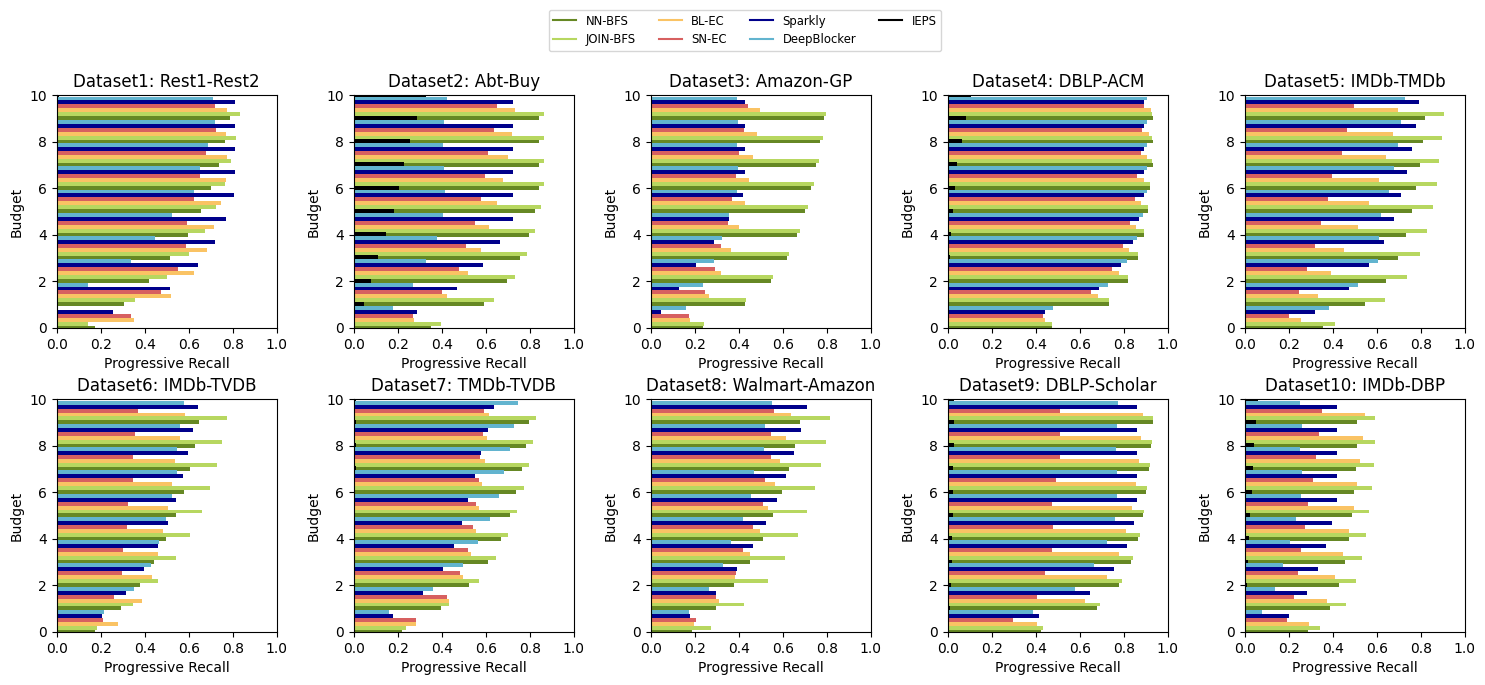}
  \vspace{-10pt}
  \caption{Progressive recall for the best progressive method per Filtering type and the baseline methods over the Record Linkage datasets.}
  \label{fig:topAUC}
\end{figure*}

\begin{figure*}[h]
  \centering
  \includegraphics[width=0.96\linewidth]{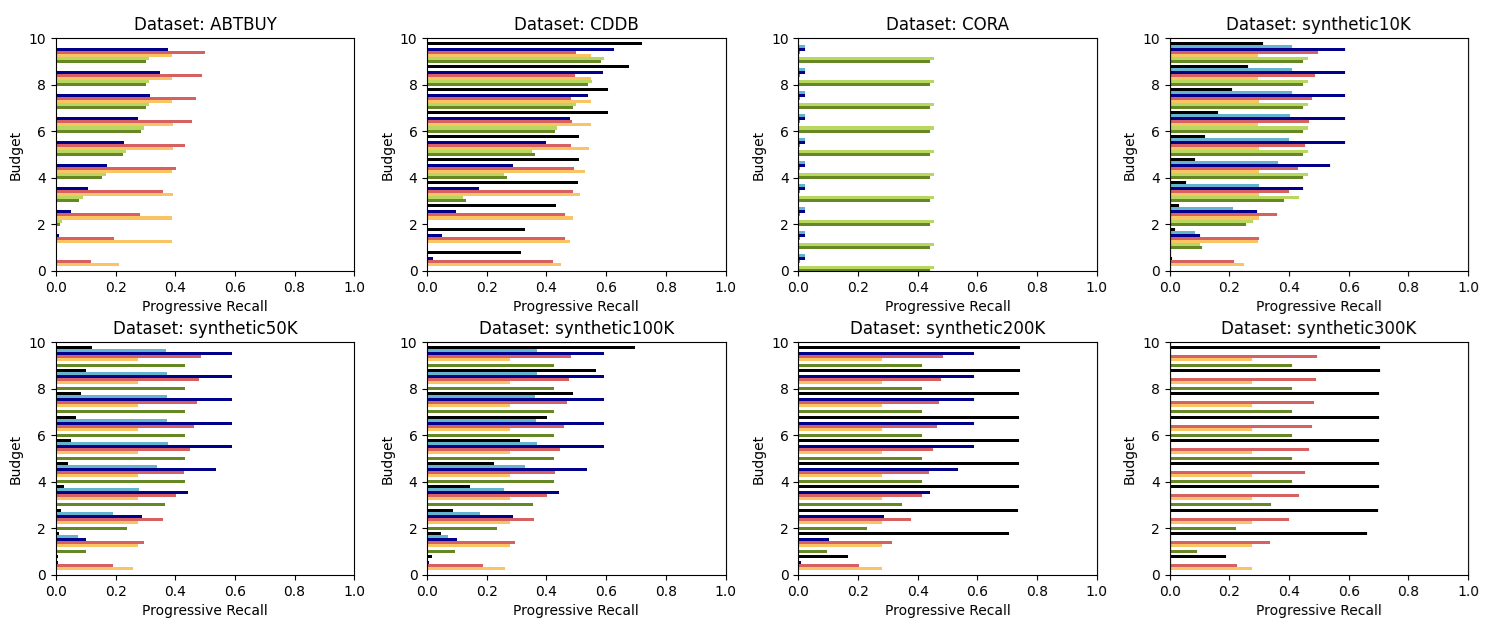}
  \vspace{-10pt}
  \caption{Progressive recall for the best progressive method per Filtering type and the baseline methods over the Deduplication datasets.}
  \label{fig:dertopAUC}
\end{figure*}

In this context, we now compare the best solutions identified in Sections \ref{sec:expNN}-\ref{sec:expSN}, i.e., the \textsf{BFS} approach of the NN and join workflows, with the \textsf{EC}  from the blocking and sorting-based workflows. Their progressive recall over the Record Linkage and Deduplication datasets is reported in Figures \ref{fig:topAUC} and \ref{fig:dertopAUC}, respectively. 

We observe two different patterns, depending on the type of dataset.
For Record Linkage,  the sorting-based solution ranks last in all datasets, but the $D_1$, with its average $DFT$ exceeding 19\% in all datasets; the larger the budget, the higher is its $DFT$, which indicates that only its top-weighted pairs are indeed duplicates. 

In contrast, the join solution outperforms all others to a statistically significant extend in seven datasets ($D_2$-$D_3$, $D_5$-$D_8$ and $D_{10}$). It is actually the top performer across all budgets in all these datasets, except for the  smallest two in $D_6$ and $D_7$. Its performance remains very high in the remaining datasets, too: in the bibliographic datasets ($D_4$ and $D_9$), its difference from the top performer (the NN solution) is statistically insignificant, while in $D_1$, it takes the lead over the blocking solution for the three largest budgets. 

The situation is reversed in most Deduplication datasets, except for the smallest one ($De_{1}$), where we observe the same patterns as in Record Linkage. In $De_{2}$ and $De_{3}$, the blocking solution takes the lead, followed in close distance by the sorting-based one, with the join workflow ranking third and fourth, respectively. However, in the five largest workflows, the sorting-based solution consistently ranks first, with a major lead over the remaining solutions. Note that the join workflow does not scale beyond 50,000 entities (i.e., $De_5$), due to insufficient memory.

The run-time measurements across all budgets are presented in the Appendix, in Figures \ref{fig:toptime} and \ref{fig:dertoptime} for the Record Linkage and Deduplication datasets, respectively. The sorting-based solution is consistently the fastest approach across all budgets, with the blocking following in close distance. The NN and join solutions are slower by 1 or 2 orders of magnitude across all datasets. In the smallest datasets, NN is slower than join, but the latter becomes much slower as the number of input entities increases. This means that the join solution exhibits poor scalability, unlike the NN one. Note that for all solutions, the size of the budget does not affect the run-time, as Filtering and Weighting are independent of the budget. Only Scheduling is affected, albeit to a minor extent.

This significant difference in the scalability of the NN and join workflows should be attributed to the relative cost of their vectorization, indexing and querying phases. For the former, vectorization is typically a time-consuming process, due to the high dimensionality and the high number of parameters of S-GTR-T5 \cite{DBLP:journals/pvldb/ZeakisPSK23}. In contrast, indexing and querying is quite efficient, due to FAISS, a the state-of-the-art tool for approximate nearest neighbor search \cite{DBLP:journals/is/AumullerBF20}. In contrast, the join workflows involve very efficient vectorization and indexing phases, with the querying one constituting the bottleneck, as it aggregates the posting lists of all tokens associated with each query entity; the number of these tokens is high, due to the schema-agnostic settings, which consider all attribute values.

The memory footprints are presented in Figures~\ref{fig:topmemory} and \ref{fig:dertopmemory} of the Appendix for the Record Linkage and theDeduplication datasets, respectively. In most cases, the memory footprint of the NN workflows is higher by an order of magnitude than the other Filtering types, because the former leverages very high dimensional embedding vectors, while the latter operate directly on string values (hence, they depend heavily on the dataset size). Note, though, that the join solution does not scale to more than 50,000, due to a two-dimensional matrix that lies at the core of its implementation that depends on the size of the input dataset.

Overall, we can conclude that for the Record Linkage datasets, the top join workflow constitutes the best solution, trading the highest accuracy for the lowest scalability in terms time and memory efficiency. In contrast, the best NN workflow favors scalability over accuracy. For the Deduplication datasets, the best sorting-based workflow outperforms all others in terms of accuracy, run-time and memory efficiency.

\subsection{Comparison to the state-of-the-art}
\label{sec:sota}

To assess the performance of the join and NN workflows, we consider two recent state-of-the-art, open-source filtering approaches: 
\begin{enumerate}[leftmargin=*]
    \item DeepBlocker\footnote{DeepBlocker is available at: \url{https://github.com/qcri/DeepBlocker}.} \cite{DBLP:conf/sigmod/MudgalLRDPKDAR18} 
    is an NN workflow that combines FAISS for indexing and querying with self-supervised learning for improving the pre-trained embedding vectors provided by FAISS. 
    \item Sparkly\footnote{Sparkly is available at: \url{https://github.com/anhaidgroup/sparkly}.} \cite{DBLP:journals/pvldb/PaulsenGD23} 
    is a join workflow that combines TF-IDF weights and BM25 scores with parallelization on top of Apache Spark. 
\end{enumerate}
For their fine-tuning, we performed grid search over the two common parameters (see Table \ref{tb:configurations}):
\begin{enumerate}[leftmargin=*]
    \item the number of nearest neighbors $k \in [1,5,10]$
    \item the indexing scheme, which can be $\{smallest, largest, both\}$
\end{enumerate}
For both methods, the best performance, which minimizes the average $DFT$, across all budgets and datasets, corresponds to $k=5$. Regarding the indexing scheme, DeepBlocker works best when indexing the smallest dataset and querying with the largest one and vice versa for Sparkly. Note that DeepBlocker uses the cosine similarity by default and that its operation is stochastic, due to the random selection of instances to be labelled to form the automatically-created training set during self-supervised learning. As a result, in each dataset and budget, we consider its average performance over 5 iterations. Note also that Sparkly is combined with the character 3-grams that optimize its performance, as shown in \cite{DBLP:journals/pvldb/PaulsenGD23}.

As an additional baseline approach, we use the best workflow generated by the Progressive Incremental Entity Resolution framework, namely I-PES \cite{DBLP:conf/edbt/GazzarriH23}. All parameters were fine-tuned according to the experimental analysis in \cite{DBLP:conf/edbt/GazzarriH23}. The available implementation, though, estimates only recall, not the progressive one.

These three baseline methods are compared with the best join and sorting-based solutions with respect to Progressive Recall over the Record Linkage and the Deduplication datasets in Figures \ref{fig:topAUC} and \ref{fig:dertopAUC}, respectively. Starting with the former, we observe that the join solution achieves the highest progressive recall for practically all budgets in all datasets, but the smallest one, where Sparkly takes the lead. In all other datasets, Sparkly and DeepBlocker underperform the join solution to a significant extent, with an average $DFT$ usually higher than 20\%. Even the sorting-based solution outperforms both Sparkly and DeepBlocker in $D_3$. 

\begin{figure*}[t]
  \centering
  \includegraphics[width=0.99\linewidth]{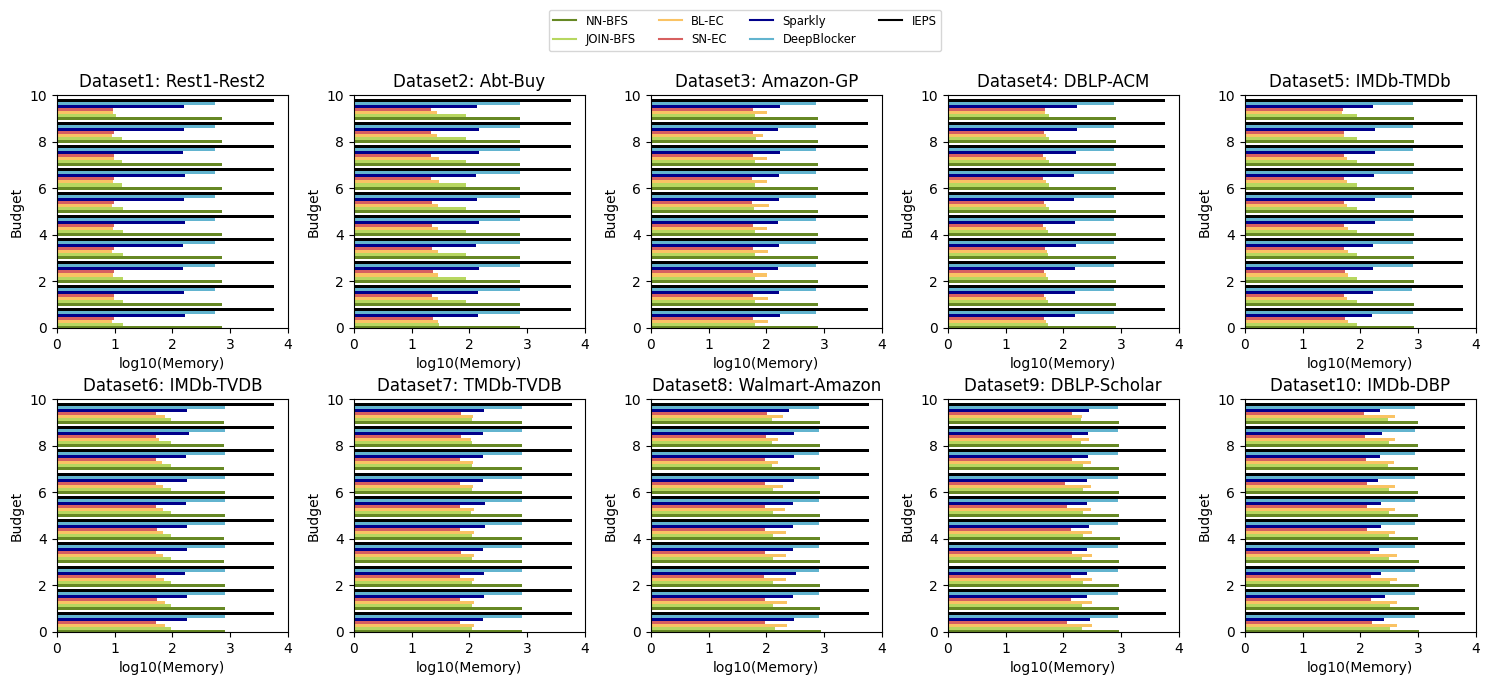}
  \vspace{-10pt}
  \caption{The memory consumption for the best approaches per Filtering type as well as for the baseline methods over the  Record Linkage datasets.}
  \label{fig:topmemory}
  \includegraphics[width=0.99\linewidth]{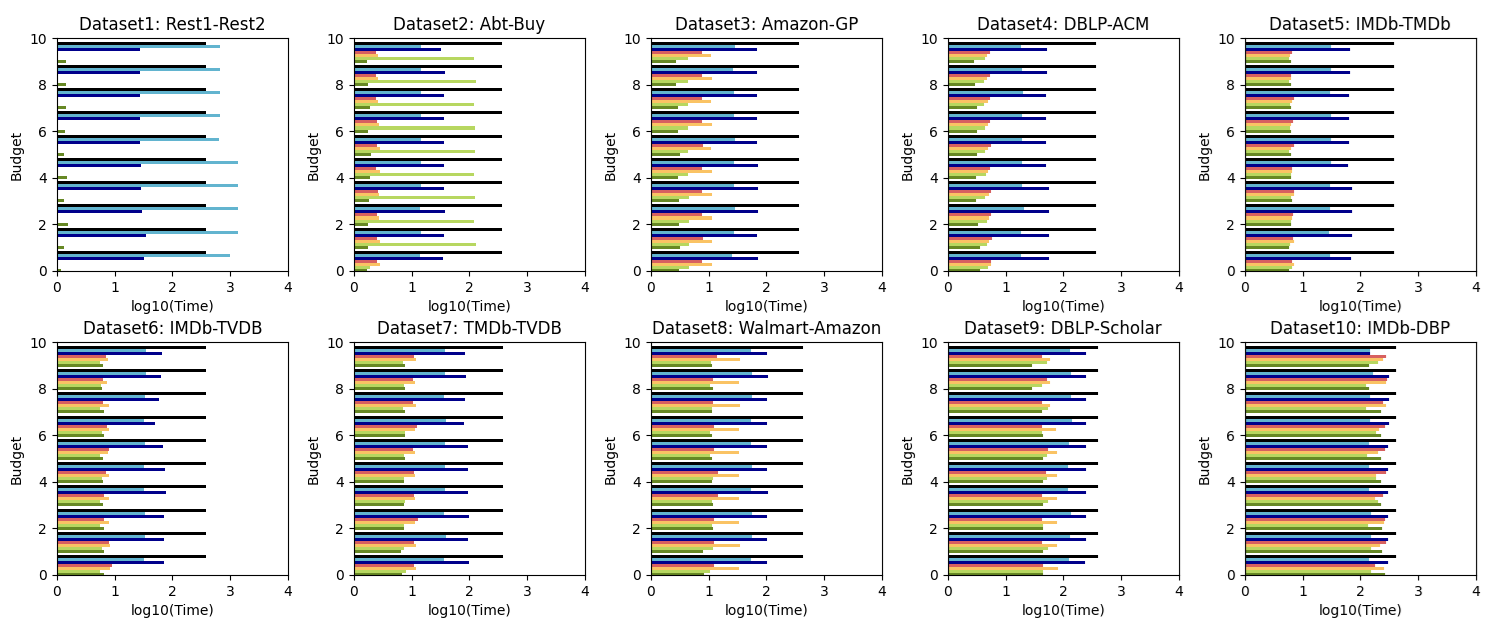}
  \vspace{-10pt}
  \caption{The run-time for the best approaches per Filtering type as well as for the baseline methods over the  Record Linkage datasets.}
  \label{fig:toptime}
\end{figure*}

\begin{figure*}[t]
  \centering
  \includegraphics[width=0.99\linewidth]{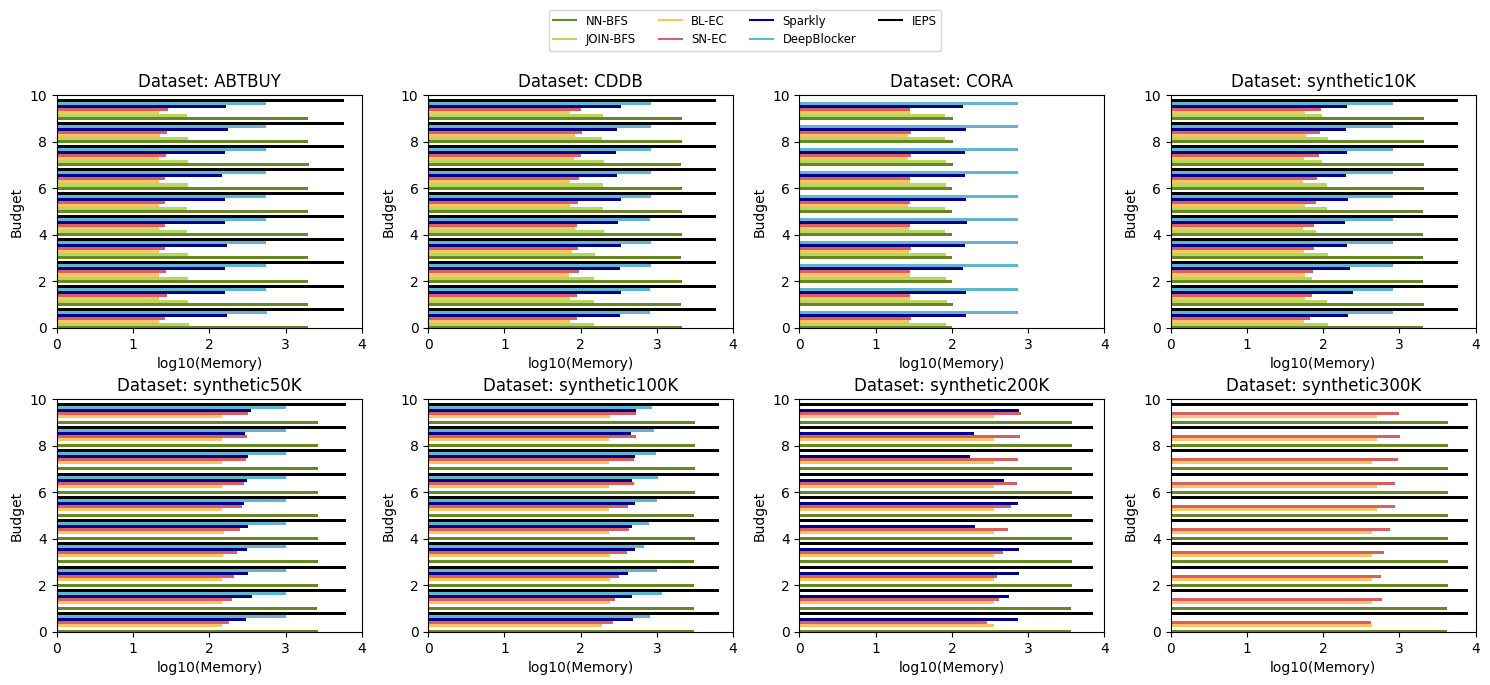}
  \vspace{-10pt}
  \caption{The memory consumption for the best approaches per Filtering type as well as for the baseline methods over the  Deduplication datasets.}
  \label{fig:dertopmemory}
  \includegraphics[width=0.99\linewidth]{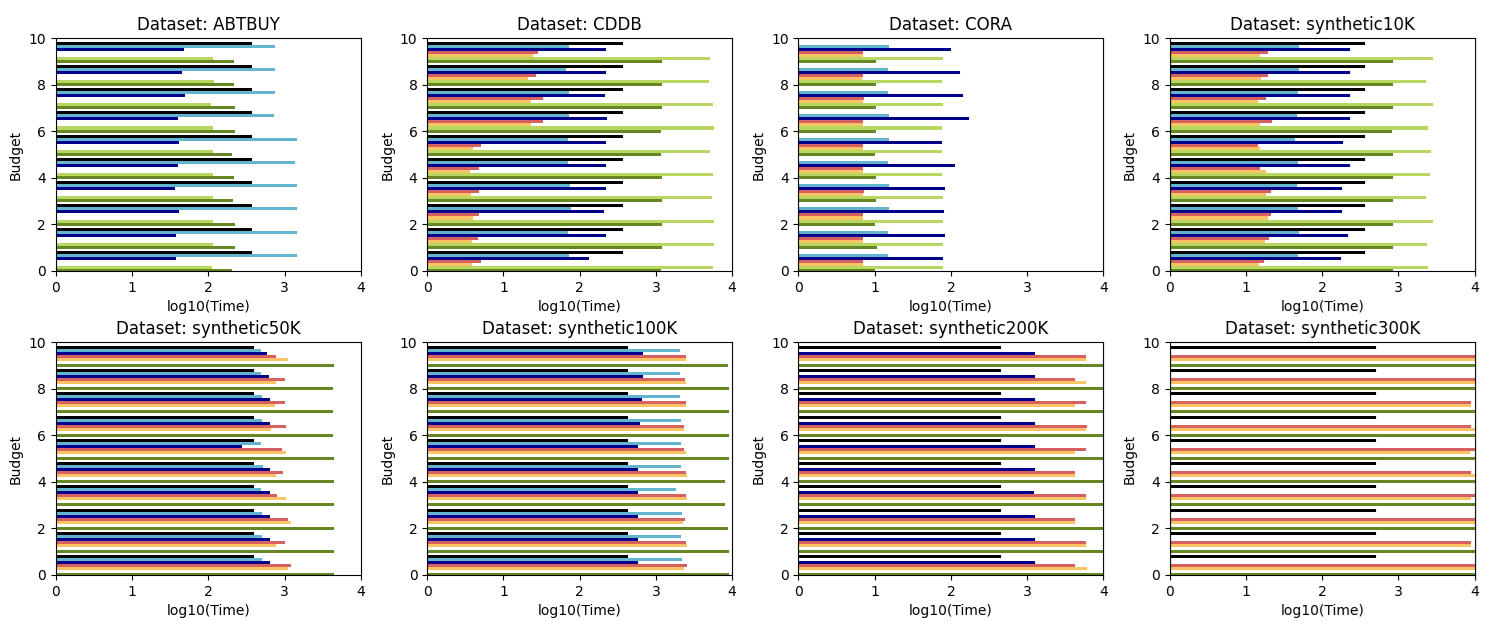}
  \vspace{-10pt}
  \caption{The run-time for the best approaches per Filtering type as well as for the baseline methods over the Deduplication datasets.}
  \label{fig:dertoptime}
\end{figure*}

In the Deduplication datasets, our techniques  outperform again the baseline methods to a significant extent. In the smallest dataset ($De_1$), the join workflow is the top performer, while in all others, the sorting-based solution takes the lead. Note that DeepBlocker scales up to $De_6$, which involves 100,000 entities, due to a two-dimensional matrix that lies at its core (similar to the join solution). The memory efficiency of these methods is presented in Figures \ref{fig:topmemory} and \ref{fig:dertopmemory}. We observe that the sorting-based workflow consistently exhibits the lowest memory footprint, even by a whole order of magnitude, especially when compared to DeepBlocker. The join solution is more memory efficient than the baseline methods in all Record Linkage datasets, where it achieves the highest accuracy.

For what concerns time efficiency, the join (and the sorting-based) solutions are consistently faster than the baseline methods over the Record Linkage datasets (Figure~\ref{fig:toptime}). Over the Deduplication datasets (Figure \ref{fig:dertoptime}), the sorting-based solution is by far the fastest up to dataset $De_4$, which contains 10,000 entities. In larger datasets, the overhead of PySpark parallelization pays off and, thus, Sparkly becomes the fastest approach.

Overall, our solutions consistently outperform the baselines with respect to effectiveness in practically all budgets of all considered datasets (except for $D_1$, where Sparkly takes the lead). They are also quite time and memory efficient, but Sparkly excels in scalability, based on its PySpark implementation.

\section{Conclusions}

We presented an architecture template for generating a wide diversity of Progressive Entity Resolution solutions based on three modules that precede the matching and the clustering algorithms in an end-to-end ER pipeline. Through a thorough experimental analysis, we identified the four top performing solutions, one for each combination of filtering and weighting techniques. 
Our solutions consistently outperform the current state-of-the-art in terms of progressive recall, memory footprint and often run-time.

In the future, we will adapt all four Filtering types to real-time ER, where the goal is to match a query record in sub-second time \cite{DBLP:conf/pakdd/RamadanC15}. For the NN and join workflows, this is a natural setting that requires fine-tuning their index. Pre-calculated similarities can be used for blocking workflows when the query record has already been indexed~\cite{DBLP:conf/cikm/ChristenGH09}. If not, Algorithm~\ref{workflow:block-1} should be adapted to work with the three indexes of DySimII, whose record insertion and query times remain practically stable, despite the increasing number of entities~\cite{DBLP:conf/pakdd/RamadanCLGH13}. Finally, Algorithm \ref{workflow:sn-1} should integrate the braided AVL trees used by F-DySNI \cite{DBLP:journals/jdiq/RamadanCLG15}, which optimize the record insertion and query times of the sorting-based workflows.

\bibliographystyle{ACM-Reference-Format}
\bibliography{references}


\begin{thebibliography}{56}


\ifx \showCODEN    \undefined \def \showCODEN     #1{\unskip}     \fi
\ifx \showDOI      \undefined \def \showDOI       #1{#1}\fi
\ifx \showISBNx    \undefined \def \showISBNx     #1{\unskip}     \fi
\ifx \showISBNxiii \undefined \def \showISBNxiii  #1{\unskip}     \fi
\ifx \showISSN     \undefined \def \showISSN      #1{\unskip}     \fi
\ifx \showLCCN     \undefined \def \showLCCN      #1{\unskip}     \fi
\ifx \shownote     \undefined \def \shownote      #1{#1}          \fi
\ifx \showarticletitle \undefined \def \showarticletitle #1{#1}   \fi
\ifx \showURL      \undefined \def \showURL       {\relax}        \fi
\providecommand\bibfield[2]{#2}
\providecommand\bibinfo[2]{#2}
\providecommand\natexlab[1]{#1}
\providecommand\showeprint[2][]{arXiv:#2}

\bibitem[Aum{\"{u}}ller et~al\mbox{.}(2020)]%
        {DBLP:journals/is/AumullerBF20}
\bibfield{author}{\bibinfo{person}{Martin Aum{\"{u}}ller}, \bibinfo{person}{Erik Bernhardsson}, {and} \bibinfo{person}{Alexander~John Faithfull}.} \bibinfo{year}{2020}\natexlab{}.
\newblock \showarticletitle{ANN-Benchmarks: {A} benchmarking tool for approximate nearest neighbor algorithms}.
\newblock \bibinfo{journal}{\emph{Inf. Syst.}}  \bibinfo{volume}{87} (\bibinfo{year}{2020}).
\newblock


\bibitem[Bojanowski et~al\mbox{.}(2016)]%
        {DBLP:journals/corr/BojanowskiGJM16}
\bibfield{author}{\bibinfo{person}{Piotr Bojanowski}, \bibinfo{person}{Edouard Grave}, \bibinfo{person}{Armand Joulin}, {and} \bibinfo{person}{Tom{\'{a}}s Mikolov}.} \bibinfo{year}{2016}\natexlab{}.
\newblock \showarticletitle{Enriching Word Vectors with Subword Information}.
\newblock \bibinfo{journal}{\emph{CoRR}}  \bibinfo{volume}{abs/1607.04606} (\bibinfo{year}{2016}).
\newblock


\bibitem[Christen(2012)]%
        {DBLP:books/daglib/0030287}
\bibfield{author}{\bibinfo{person}{Peter Christen}.} \bibinfo{year}{2012}\natexlab{}.
\newblock \bibinfo{booktitle}{\emph{Data Matching - Concepts and Techniques for Record Linkage, Entity Resolution, and Duplicate Detection}}.
\newblock \bibinfo{publisher}{Springer}.
\newblock


\bibitem[Christen et~al\mbox{.}(2009)]%
        {DBLP:conf/cikm/ChristenGH09}
\bibfield{author}{\bibinfo{person}{Peter Christen}, \bibinfo{person}{Ross~W. Gayler}, {and} \bibinfo{person}{David Hawking}.} \bibinfo{year}{2009}\natexlab{}.
\newblock \showarticletitle{Similarity-aware indexing for real-time entity resolution}. In \bibinfo{booktitle}{\emph{{CIKM}}}. \bibinfo{pages}{1565--1568}.
\newblock


\bibitem[Christophides et~al\mbox{.}(2021)]%
        {DBLP:journals/csur/ChristophidesEP21}
\bibfield{author}{\bibinfo{person}{Vassilis Christophides}, \bibinfo{person}{Vasilis Efthymiou}, \bibinfo{person}{Themis Palpanas}, \bibinfo{person}{George Papadakis}, {and} \bibinfo{person}{Kostas Stefanidis}.} \bibinfo{year}{2021}\natexlab{}.
\newblock \showarticletitle{An Overview of End-to-End Entity Resolution for Big Data}.
\newblock \bibinfo{journal}{\emph{{ACM} Comput. Surv.}} \bibinfo{volume}{53}, \bibinfo{number}{6} (\bibinfo{year}{2021}), \bibinfo{pages}{127:1--127:42}.
\newblock


\bibitem[Devlin et~al\mbox{.}(2019)]%
        {DBLP:conf/naacl/DevlinCLT19}
\bibfield{author}{\bibinfo{person}{Jacob Devlin}, \bibinfo{person}{Ming{-}Wei Chang}, \bibinfo{person}{Kenton Lee}, {and} \bibinfo{person}{Kristina Toutanova}.} \bibinfo{year}{2019}\natexlab{}.
\newblock \showarticletitle{{BERT:} Pre-training of Deep Bidirectional Transformers for Language Understanding}. In \bibinfo{booktitle}{\emph{{NAACL-HLT}}}. \bibinfo{pages}{4171--4186}.
\newblock


\bibitem[Dong and Srivastava(2015)]%
        {DBLP:series/synthesis/2015Dong}
\bibfield{author}{\bibinfo{person}{Xin~Luna Dong} {and} \bibinfo{person}{Divesh Srivastava}.} \bibinfo{year}{2015}\natexlab{}.
\newblock \bibinfo{booktitle}{\emph{Big Data Integration}}.
\newblock \bibinfo{publisher}{Morgan {\&} Claypool Publishers}.
\newblock


\bibitem[Fan et~al\mbox{.}(2024)]%
        {DBLP:journals/sigmod/FanTLWDJGT24}
\bibfield{author}{\bibinfo{person}{Ju Fan}, \bibinfo{person}{Jianhong Tu}, \bibinfo{person}{Guoliang Li}, \bibinfo{person}{Peng Wang}, \bibinfo{person}{Xiaoyong Du}, \bibinfo{person}{Xiaofeng Jia}, \bibinfo{person}{Song Gao}, {and} \bibinfo{person}{Nan Tang}.} \bibinfo{year}{2024}\natexlab{}.
\newblock \showarticletitle{Unicorn: {A} Unified Multi-Tasking Matching Model}.
\newblock \bibinfo{journal}{\emph{{SIGMOD} Rec.}} \bibinfo{volume}{53}, \bibinfo{number}{1} (\bibinfo{year}{2024}), \bibinfo{pages}{44--53}.
\newblock


\bibitem[Galhotra et~al\mbox{.}(2021a)]%
        {DBLP:conf/sigmod/GalhotraFSS21}
\bibfield{author}{\bibinfo{person}{Sainyam Galhotra}, \bibinfo{person}{Donatella Firmani}, \bibinfo{person}{Barna Saha}, {and} \bibinfo{person}{Divesh Srivastava}.} \bibinfo{year}{2021}\natexlab{a}.
\newblock \showarticletitle{{BEER:} Blocking for Effective Entity Resolution}. In \bibinfo{booktitle}{\emph{{SIGMOD}}}. \bibinfo{pages}{2711--2715}.
\newblock


\bibitem[Galhotra et~al\mbox{.}(2021b)]%
        {DBLP:journals/vldb/GalhotraFSS21}
\bibfield{author}{\bibinfo{person}{Sainyam Galhotra}, \bibinfo{person}{Donatella Firmani}, \bibinfo{person}{Barna Saha}, {and} \bibinfo{person}{Divesh Srivastava}.} \bibinfo{year}{2021}\natexlab{b}.
\newblock \showarticletitle{Efficient and effective {ER} with progressive blocking}.
\newblock \bibinfo{journal}{\emph{{VLDB} J.}} \bibinfo{volume}{30}, \bibinfo{number}{4} (\bibinfo{year}{2021}), \bibinfo{pages}{537--557}.
\newblock


\bibitem[Gazzarri and Herschel(2023)]%
        {DBLP:conf/edbt/GazzarriH23}
\bibfield{author}{\bibinfo{person}{Leonardo Gazzarri} {and} \bibinfo{person}{Melanie Herschel}.} \bibinfo{year}{2023}\natexlab{}.
\newblock \showarticletitle{Progressive Entity Resolution over Incremental Data}. In \bibinfo{booktitle}{\emph{{EDBT}}}. \bibinfo{pages}{80--91}.
\newblock


\bibitem[Genossar et~al\mbox{.}(2023)]%
        {DBLP:journals/pacmmod/GenossarGS23}
\bibfield{author}{\bibinfo{person}{Bar Genossar}, \bibinfo{person}{Avigdor Gal}, {and} \bibinfo{person}{Roee Shraga}.} \bibinfo{year}{2023}\natexlab{}.
\newblock \showarticletitle{The Battleship Approach to the Low Resource Entity Matching Problem}.
\newblock \bibinfo{journal}{\emph{Proc. {ACM} Manag. Data}} \bibinfo{volume}{1}, \bibinfo{number}{4} (\bibinfo{year}{2023}), \bibinfo{pages}{224:1--224:25}.
\newblock


\bibitem[Getoor and Machanavajjhala(2012)]%
        {DBLP:journals/pvldb/GetoorM12}
\bibfield{author}{\bibinfo{person}{Lise Getoor} {and} \bibinfo{person}{Ashwin Machanavajjhala}.} \bibinfo{year}{2012}\natexlab{}.
\newblock \showarticletitle{Entity Resolution: Theory, Practice {\&} Open Challenges}.
\newblock \bibinfo{journal}{\emph{PVLDB}} \bibinfo{volume}{5}, \bibinfo{number}{12} (\bibinfo{year}{2012}), \bibinfo{pages}{2018--2019}.
\newblock


\bibitem[Hassanzadeh et~al\mbox{.}(2009)]%
        {DBLP:journals/pvldb/HassanzadehCML09}
\bibfield{author}{\bibinfo{person}{Oktie Hassanzadeh}, \bibinfo{person}{Fei Chiang}, \bibinfo{person}{Ren{\'{e}}e~J. Miller}, {and} \bibinfo{person}{Hyun~Chul Lee}.} \bibinfo{year}{2009}\natexlab{}.
\newblock \showarticletitle{Framework for Evaluating Clustering Algorithms in Duplicate Detection}.
\newblock \bibinfo{journal}{\emph{PVLDB}} \bibinfo{volume}{2}, \bibinfo{number}{1} (\bibinfo{year}{2009}), \bibinfo{pages}{1282--1293}.
\newblock


\bibitem[Jain et~al\mbox{.}(2021)]%
        {DBLP:journals/pvldb/JainSS21}
\bibfield{author}{\bibinfo{person}{Arjit Jain}, \bibinfo{person}{Sunita Sarawagi}, {and} \bibinfo{person}{Prithviraj Sen}.} \bibinfo{year}{2021}\natexlab{}.
\newblock \showarticletitle{Deep Indexed Active Learning for Matching Heterogeneous Entity Representations}.
\newblock \bibinfo{journal}{\emph{PVLDB}} \bibinfo{volume}{15}, \bibinfo{number}{1} (\bibinfo{year}{2021}), \bibinfo{pages}{31--45}.
\newblock


\bibitem[Johnson et~al\mbox{.}(2021)]%
        {DBLP:journals/tbd/JohnsonDJ21}
\bibfield{author}{\bibinfo{person}{Jeff Johnson}, \bibinfo{person}{Matthijs Douze}, {and} \bibinfo{person}{Herv{\'{e}} J{\'{e}}gou}.} \bibinfo{year}{2021}\natexlab{}.
\newblock \showarticletitle{Billion-Scale Similarity Search with GPUs}.
\newblock \bibinfo{journal}{\emph{{IEEE} Trans. Big Data}} \bibinfo{volume}{7}, \bibinfo{number}{3} (\bibinfo{year}{2021}), \bibinfo{pages}{535--547}.
\newblock


\bibitem[K{\"{o}}pcke et~al\mbox{.}(2010)]%
        {DBLP:journals/pvldb/KopckeTR10}
\bibfield{author}{\bibinfo{person}{Hanna K{\"{o}}pcke}, \bibinfo{person}{Andreas Thor}, {and} \bibinfo{person}{Erhard Rahm}.} \bibinfo{year}{2010}\natexlab{}.
\newblock \showarticletitle{Evaluation of entity resolution approaches on real-world match problems}.
\newblock \bibinfo{journal}{\emph{PVLDB}} \bibinfo{volume}{3}, \bibinfo{number}{1} (\bibinfo{year}{2010}), \bibinfo{pages}{484--493}.
\newblock


\bibitem[Lan et~al\mbox{.}(2020)]%
        {DBLP:conf/iclr/LanCGGSS20}
\bibfield{author}{\bibinfo{person}{Zhenzhong Lan}, \bibinfo{person}{Mingda Chen}, \bibinfo{person}{Sebastian Goodman}, \bibinfo{person}{Kevin Gimpel}, \bibinfo{person}{Piyush Sharma}, {and} \bibinfo{person}{Radu Soricut}.} \bibinfo{year}{2020}\natexlab{}.
\newblock \showarticletitle{{ALBERT:} {A} Lite {BERT} for Self-supervised Learning of Language Representations}. In \bibinfo{booktitle}{\emph{{ICLR}}}.
\newblock


\bibitem[Liu et~al\mbox{.}(2019)]%
        {DBLP:journals/corr/abs-1907-11692}
\bibfield{author}{\bibinfo{person}{Yinhan Liu}, \bibinfo{person}{Myle Ott}, \bibinfo{person}{Naman Goyal}, \bibinfo{person}{Jingfei Du}, \bibinfo{person}{Mandar Joshi}, \bibinfo{person}{Danqi Chen}, \bibinfo{person}{Omer Levy}, \bibinfo{person}{Mike Lewis}, \bibinfo{person}{Luke Zettlemoyer}, {and} \bibinfo{person}{Veselin Stoyanov}.} \bibinfo{year}{2019}\natexlab{}.
\newblock \showarticletitle{RoBERTa: {A} Robustly Optimized {BERT} Pretraining Approach}.
\newblock \bibinfo{journal}{\emph{CoRR}}  \bibinfo{volume}{abs/1907.11692} (\bibinfo{year}{2019}).
\newblock


\bibitem[Meduri et~al\mbox{.}(2020)]%
        {DBLP:conf/sigmod/Meduri0SS20}
\bibfield{author}{\bibinfo{person}{Venkata~Vamsikrishna Meduri}, \bibinfo{person}{Lucian Popa}, \bibinfo{person}{Prithviraj Sen}, {and} \bibinfo{person}{Mohamed Sarwat}.} \bibinfo{year}{2020}\natexlab{}.
\newblock \showarticletitle{A Comprehensive Benchmark Framework for Active Learning Methods in Entity Matching}. In \bibinfo{booktitle}{\emph{{SIGMOD}}}. \bibinfo{pages}{1133--1147}.
\newblock


\bibitem[Mikolov et~al\mbox{.}(2013a)]%
        {DBLP:journals/corr/abs-1301-3781}
\bibfield{author}{\bibinfo{person}{Tom{\'{a}}s Mikolov}, \bibinfo{person}{Kai Chen}, \bibinfo{person}{Greg Corrado}, {and} \bibinfo{person}{Jeffrey Dean}.} \bibinfo{year}{2013}\natexlab{a}.
\newblock \showarticletitle{Efficient Estimation of Word Representations in Vector Space}. In \bibinfo{booktitle}{\emph{{ICLR}}}.
\newblock


\bibitem[Mikolov et~al\mbox{.}(2013b)]%
        {DBLP:conf/nips/MikolovSCCD13}
\bibfield{author}{\bibinfo{person}{Tom{\'{a}}s Mikolov}, \bibinfo{person}{Ilya Sutskever}, \bibinfo{person}{Kai Chen}, \bibinfo{person}{Gregory~S. Corrado}, {and} \bibinfo{person}{Jeffrey Dean}.} \bibinfo{year}{2013}\natexlab{b}.
\newblock \showarticletitle{Distributed Representations of Words and Phrases and their Compositionality}. In \bibinfo{booktitle}{\emph{NIPS}}. \bibinfo{pages}{3111--3119}.
\newblock


\bibitem[Mudgal et~al\mbox{.}(2018)]%
        {DBLP:conf/sigmod/MudgalLRDPKDAR18}
\bibfield{author}{\bibinfo{person}{Sidharth Mudgal}, \bibinfo{person}{Han Li}, \bibinfo{person}{Theodoros Rekatsinas}, \bibinfo{person}{AnHai Doan}, \bibinfo{person}{Youngchoon Park}, \bibinfo{person}{Ganesh Krishnan}, \bibinfo{person}{Rohit Deep}, \bibinfo{person}{Esteban Arcaute}, {and} \bibinfo{person}{Vijay Raghavendra}.} \bibinfo{year}{2018}\natexlab{}.
\newblock \showarticletitle{Deep Learning for Entity Matching: {A} Design Space Exploration}. In \bibinfo{booktitle}{\emph{{SIGMOD}}}. \bibinfo{publisher}{{ACM}}, \bibinfo{pages}{19--34}.
\newblock


\bibitem[Neuhof et~al\mbox{.}(2024)]%
        {neuhof2024open}
\bibfield{author}{\bibinfo{person}{Franziska Neuhof}, \bibinfo{person}{Marco Fisichella}, \bibinfo{person}{George Papadakis}, \bibinfo{person}{Konstantinos Nikoletos}, \bibinfo{person}{Nikolaus Augsten}, \bibinfo{person}{Wolfgang Nejdl}, {and} \bibinfo{person}{Manolis Koubarakis}.} \bibinfo{year}{2024}\natexlab{}.
\newblock \showarticletitle{Open benchmark for filtering techniques in entity resolution}.
\newblock \bibinfo{journal}{\emph{The VLDB Journal}} (\bibinfo{year}{2024}), \bibinfo{pages}{1--26}.
\newblock


\bibitem[Obraczka et~al\mbox{.}(2021)]%
        {DBLP:conf/esws/ObraczkaSR21}
\bibfield{author}{\bibinfo{person}{Daniel Obraczka}, \bibinfo{person}{Jonathan Schuchart}, {and} \bibinfo{person}{Erhard Rahm}.} \bibinfo{year}{2021}\natexlab{}.
\newblock \showarticletitle{Embedding-Assisted Entity Resolution for Knowledge Graphs}. In \bibinfo{booktitle}{\emph{{KGCW}@ESWC}}.
\newblock


\bibitem[Papadakis et~al\mbox{.}(2023a)]%
        {DBLP:journals/vldb/PapadakisETHC23}
\bibfield{author}{\bibinfo{person}{George Papadakis}, \bibinfo{person}{Vasilis Efthymiou}, \bibinfo{person}{Emmanouil Thanos}, \bibinfo{person}{Oktie Hassanzadeh}, {and} \bibinfo{person}{Peter Christen}.} \bibinfo{year}{2023}\natexlab{a}.
\newblock \showarticletitle{An analysis of one-to-one matching algorithms for entity resolution}.
\newblock \bibinfo{journal}{\emph{{VLDB} J.}} \bibinfo{volume}{32}, \bibinfo{number}{6} (\bibinfo{year}{2023}), \bibinfo{pages}{1369--1400}.
\newblock


\bibitem[Papadakis et~al\mbox{.}(2023b)]%
        {DBLP:conf/icde/0001FSMAN23}
\bibfield{author}{\bibinfo{person}{George Papadakis}, \bibinfo{person}{Marco Fisichella}, \bibinfo{person}{Franziska Schoger}, \bibinfo{person}{George Mandilaras}, \bibinfo{person}{Nikolaus Augsten}, {and} \bibinfo{person}{Wolfgang Nejdl}.} \bibinfo{year}{2023}\natexlab{b}.
\newblock \showarticletitle{Benchmarking Filtering Techniques for Entity Resolution}. In \bibinfo{booktitle}{\emph{{ICDE}}}. \bibinfo{publisher}{{IEEE}}, \bibinfo{pages}{653--666}.
\newblock


\bibitem[Papadakis et~al\mbox{.}(2021a)]%
        {DBLP:series/synthesis/2021Papadakis}
\bibfield{author}{\bibinfo{person}{George Papadakis}, \bibinfo{person}{Ekaterini Ioannou}, \bibinfo{person}{Emanouil Thanos}, {and} \bibinfo{person}{Themis Palpanas}.} \bibinfo{year}{2021}\natexlab{a}.
\newblock \bibinfo{booktitle}{\emph{The Four Generations of Entity Resolution}}.
\newblock \bibinfo{publisher}{Morgan {\&} Claypool Publishers}.
\newblock


\bibitem[Papadakis et~al\mbox{.}(2024)]%
        {DBLP:conf/icde/0001KCP24}
\bibfield{author}{\bibinfo{person}{George Papadakis}, \bibinfo{person}{Nishadi Kirielle}, \bibinfo{person}{Peter Christen}, {and} \bibinfo{person}{Themis Palpanas}.} \bibinfo{year}{2024}\natexlab{}.
\newblock \showarticletitle{A Critical Re-evaluation of Record Linkage Benchmarks for Learning-Based Matching Algorithms}. In \bibinfo{booktitle}{\emph{{ICDE}}}. \bibinfo{pages}{3435--3448}.
\newblock


\bibitem[Papadakis et~al\mbox{.}(2020)]%
        {DBLP:journals/is/PapadakisMGSTGB20}
\bibfield{author}{\bibinfo{person}{George Papadakis}, \bibinfo{person}{Georgios~M. Mandilaras}, \bibinfo{person}{Luca Gagliardelli}, \bibinfo{person}{Giovanni Simonini}, \bibinfo{person}{Emmanouil Thanos}, \bibinfo{person}{George Giannakopoulos}, \bibinfo{person}{Sonia Bergamaschi}, \bibinfo{person}{Themis Palpanas}, {and} \bibinfo{person}{Manolis Koubarakis}.} \bibinfo{year}{2020}\natexlab{}.
\newblock \showarticletitle{Three-dimensional Entity Resolution with JedAI}.
\newblock \bibinfo{journal}{\emph{Inf. Syst.}}  \bibinfo{volume}{93} (\bibinfo{year}{2020}), \bibinfo{pages}{101565}.
\newblock


\bibitem[Papadakis et~al\mbox{.}(2021b)]%
        {DBLP:journals/csur/PapadakisSTP20}
\bibfield{author}{\bibinfo{person}{George Papadakis}, \bibinfo{person}{Dimitrios Skoutas}, \bibinfo{person}{Emmanouil Thanos}, {and} \bibinfo{person}{Themis Palpanas}.} \bibinfo{year}{2021}\natexlab{b}.
\newblock \showarticletitle{Blocking and Filtering Techniques for Entity Resolution: {A} Survey}.
\newblock \bibinfo{journal}{\emph{{ACM} Comput. Surv.}} \bibinfo{volume}{53}, \bibinfo{number}{2} (\bibinfo{year}{2021}), \bibinfo{pages}{31:1--31:42}.
\newblock


\bibitem[Papadakis et~al\mbox{.}(2016)]%
        {DBLP:journals/pvldb/0001SGP16}
\bibfield{author}{\bibinfo{person}{George Papadakis}, \bibinfo{person}{Jonathan Svirsky}, \bibinfo{person}{Avigdor Gal}, {and} \bibinfo{person}{Themis Palpanas}.} \bibinfo{year}{2016}\natexlab{}.
\newblock \showarticletitle{Comparative Analysis of Approximate Blocking Techniques for Entity Resolution}.
\newblock \bibinfo{journal}{\emph{PVLDB}} \bibinfo{volume}{9}, \bibinfo{number}{9} (\bibinfo{year}{2016}), \bibinfo{pages}{684--695}.
\newblock


\bibitem[Papenbrock et~al\mbox{.}(2015)]%
        {DBLP:journals/tkde/PapenbrockHN15}
\bibfield{author}{\bibinfo{person}{Thorsten Papenbrock}, \bibinfo{person}{Arvid Heise}, {and} \bibinfo{person}{Felix Naumann}.} \bibinfo{year}{2015}\natexlab{}.
\newblock \showarticletitle{Progressive Duplicate Detection}.
\newblock \bibinfo{journal}{\emph{TKDE}} \bibinfo{volume}{27}, \bibinfo{number}{5} (\bibinfo{year}{2015}), \bibinfo{pages}{1316--1329}.
\newblock


\bibitem[Paulsen et~al\mbox{.}(2023)]%
        {DBLP:journals/pvldb/PaulsenGD23}
\bibfield{author}{\bibinfo{person}{Derek Paulsen}, \bibinfo{person}{Yash Govind}, {and} \bibinfo{person}{AnHai Doan}.} \bibinfo{year}{2023}\natexlab{}.
\newblock \showarticletitle{Sparkly: {A} Simple yet Surprisingly Strong {TF/IDF} Blocker for Entity Matching}.
\newblock \bibinfo{journal}{\emph{PVLDB}} \bibinfo{volume}{16}, \bibinfo{number}{6} (\bibinfo{year}{2023}), \bibinfo{pages}{1507--1519}.
\newblock


\bibitem[Pennington et~al\mbox{.}(2014)]%
        {DBLP:conf/emnlp/PenningtonSM14}
\bibfield{author}{\bibinfo{person}{Jeffrey Pennington}, \bibinfo{person}{Richard Socher}, {and} \bibinfo{person}{Christopher~D. Manning}.} \bibinfo{year}{2014}\natexlab{}.
\newblock \showarticletitle{Glove: Global Vectors for Word Representation}. In \bibinfo{booktitle}{\emph{{EMNLP}}}. \bibinfo{pages}{1532--1543}.
\newblock


\bibitem[Raffel et~al\mbox{.}(2020)]%
        {DBLP:journals/jmlr/RaffelSRLNMZLL20}
\bibfield{author}{\bibinfo{person}{Colin Raffel}, \bibinfo{person}{Noam Shazeer}, \bibinfo{person}{Adam Roberts}, \bibinfo{person}{Katherine Lee}, \bibinfo{person}{Sharan Narang}, \bibinfo{person}{Michael Matena}, \bibinfo{person}{Yanqi Zhou}, \bibinfo{person}{Wei Li}, {and} \bibinfo{person}{Peter~J. Liu}.} \bibinfo{year}{2020}\natexlab{}.
\newblock \showarticletitle{Exploring the Limits of Transfer Learning with a Unified Text-to-Text Transformer}.
\newblock \bibinfo{journal}{\emph{J. Mach. Learn. Res.}}  \bibinfo{volume}{21} (\bibinfo{year}{2020}), \bibinfo{pages}{140:1--140:67}.
\newblock


\bibitem[Ramadan and Christen(2015)]%
        {DBLP:conf/pakdd/RamadanC15}
\bibfield{author}{\bibinfo{person}{Banda Ramadan} {and} \bibinfo{person}{Peter Christen}.} \bibinfo{year}{2015}\natexlab{}.
\newblock \showarticletitle{Unsupervised Blocking Key Selection for Real-Time Entity Resolution}. In \bibinfo{booktitle}{\emph{{PAKDD}}}. \bibinfo{pages}{574--585}.
\newblock


\bibitem[Ramadan et~al\mbox{.}(2015)]%
        {DBLP:journals/jdiq/RamadanCLG15}
\bibfield{author}{\bibinfo{person}{Banda Ramadan}, \bibinfo{person}{Peter Christen}, \bibinfo{person}{Huizhi Liang}, {and} \bibinfo{person}{Ross~W. Gayler}.} \bibinfo{year}{2015}\natexlab{}.
\newblock \showarticletitle{Dynamic Sorted Neighborhood Indexing for Real-Time Entity Resolution}.
\newblock \bibinfo{journal}{\emph{{ACM} J. Data Inf. Qual.}} \bibinfo{volume}{6}, \bibinfo{number}{4} (\bibinfo{year}{2015}), \bibinfo{pages}{15:1--15:29}.
\newblock


\bibitem[Ramadan et~al\mbox{.}(2013)]%
        {DBLP:conf/pakdd/RamadanCLGH13}
\bibfield{author}{\bibinfo{person}{Banda Ramadan}, \bibinfo{person}{Peter Christen}, \bibinfo{person}{Huizhi Liang}, \bibinfo{person}{Ross~W. Gayler}, {and} \bibinfo{person}{David Hawking}.} \bibinfo{year}{2013}\natexlab{}.
\newblock \showarticletitle{Dynamic Similarity-Aware Inverted Indexing for Real-Time Entity Resolution}. In \bibinfo{booktitle}{\emph{{PAKDD}}}. \bibinfo{pages}{47--58}.
\newblock


\bibitem[Saeedi et~al\mbox{.}(2018)]%
        {DBLP:journals/csimq/SaeediNPR18}
\bibfield{author}{\bibinfo{person}{Alieh Saeedi}, \bibinfo{person}{Markus Nentwig}, \bibinfo{person}{Eric Peukert}, {and} \bibinfo{person}{Erhard Rahm}.} \bibinfo{year}{2018}\natexlab{}.
\newblock \showarticletitle{Scalable Matching and Clustering of Entities with {FAMER}}.
\newblock \bibinfo{journal}{\emph{Complex Syst. Informatics Model. Q.}}  \bibinfo{volume}{16} (\bibinfo{year}{2018}), \bibinfo{pages}{61--83}.
\newblock


\bibitem[Sanh et~al\mbox{.}(2019)]%
        {DBLP:journals/corr/abs-1910-01108}
\bibfield{author}{\bibinfo{person}{Victor Sanh}, \bibinfo{person}{Lysandre Debut}, \bibinfo{person}{Julien Chaumond}, {and} \bibinfo{person}{Thomas Wolf}.} \bibinfo{year}{2019}\natexlab{}.
\newblock \showarticletitle{DistilBERT, a distilled version of {BERT:} smaller, faster, cheaper and lighter}.
\newblock \bibinfo{journal}{\emph{CoRR}}  \bibinfo{volume}{abs/1910.01108} (\bibinfo{year}{2019}).
\newblock


\bibitem[Sarawagi and Bhamidipaty(2002)]%
        {DBLP:conf/kdd/SarawagiB02}
\bibfield{author}{\bibinfo{person}{Sunita Sarawagi} {and} \bibinfo{person}{Anuradha Bhamidipaty}.} \bibinfo{year}{2002}\natexlab{}.
\newblock \showarticletitle{Interactive deduplication using active learning}. In \bibinfo{booktitle}{\emph{{KDD}}}. \bibinfo{publisher}{{ACM}}, \bibinfo{pages}{269--278}.
\newblock


\bibitem[Shvaiko et~al\mbox{.}(2010)]%
        {DBLP:conf/semweb/2010om}
\bibfield{editor}{\bibinfo{person}{Pavel Shvaiko}, \bibinfo{person}{J{\'{e}}r{\^{o}}me Euzenat}, \bibinfo{person}{Fausto Giunchiglia}, \bibinfo{person}{Heiner Stuckenschmidt}, \bibinfo{person}{Ming Mao}, {and} \bibinfo{person}{Isabel~F. Cruz}} (Eds.). \bibinfo{year}{2010}\natexlab{}.
\newblock \bibinfo{booktitle}{\emph{Proceedings of the 5th International Workshop on Ontology Matching (OM-2010), Shanghai, China, November 7, 2010}}. Vol.~\bibinfo{volume}{689}.
\newblock


\bibitem[Simonini et~al\mbox{.}(2019)]%
        {DBLP:journals/tkde/SimoniniPPB19}
\bibfield{author}{\bibinfo{person}{Giovanni Simonini}, \bibinfo{person}{George Papadakis}, \bibinfo{person}{Themis Palpanas}, {and} \bibinfo{person}{Sonia Bergamaschi}.} \bibinfo{year}{2019}\natexlab{}.
\newblock \showarticletitle{Schema-Agnostic Progressive Entity Resolution}.
\newblock \bibinfo{journal}{\emph{TKDE}} \bibinfo{volume}{31}, \bibinfo{number}{6} (\bibinfo{year}{2019}), \bibinfo{pages}{1208--1221}.
\newblock


\bibitem[Simonini et~al\mbox{.}(2022)]%
        {DBLP:journals/pvldb/SimoniniZBN22}
\bibfield{author}{\bibinfo{person}{Giovanni Simonini}, \bibinfo{person}{Luca Zecchini}, \bibinfo{person}{Sonia Bergamaschi}, {and} \bibinfo{person}{Felix Naumann}.} \bibinfo{year}{2022}\natexlab{}.
\newblock \showarticletitle{Entity Resolution On-Demand}.
\newblock \bibinfo{journal}{\emph{PVLDB}} \bibinfo{volume}{15}, \bibinfo{number}{7} (\bibinfo{year}{2022}), \bibinfo{pages}{1506--1518}.
\newblock


\bibitem[Song et~al\mbox{.}(2020)]%
        {DBLP:conf/nips/Song0QLL20}
\bibfield{author}{\bibinfo{person}{Kaitao Song}, \bibinfo{person}{Xu Tan}, \bibinfo{person}{Tao Qin}, \bibinfo{person}{Jianfeng Lu}, {and} \bibinfo{person}{Tie{-}Yan Liu}.} \bibinfo{year}{2020}\natexlab{}.
\newblock \showarticletitle{MPNet: Masked and Permuted Pre-training for Language Understanding}. In \bibinfo{booktitle}{\emph{NIPS}}.
\newblock


\bibitem[Sun et~al\mbox{.}(2022)]%
        {DBLP:journals/access/SunHSN22}
\bibfield{author}{\bibinfo{person}{Chenchen Sun}, \bibinfo{person}{Zhijiang Hou}, \bibinfo{person}{Derong Shen}, {and} \bibinfo{person}{Tiezheng Nie}.} \bibinfo{year}{2022}\natexlab{}.
\newblock \showarticletitle{Progressive Entity Matching via Cost Benefit Analysis}.
\newblock \bibinfo{journal}{\emph{{IEEE} Access}}  \bibinfo{volume}{10} (\bibinfo{year}{2022}), \bibinfo{pages}{3979--3989}.
\newblock


\bibitem[Sun et~al\mbox{.}(2023)]%
        {DBLP:conf/adma/SunJXSNW23}
\bibfield{author}{\bibinfo{person}{Chenchen Sun}, \bibinfo{person}{Yuyuan Jin}, \bibinfo{person}{Yang Xu}, \bibinfo{person}{Derong Shen}, \bibinfo{person}{Tiezheng Nie}, {and} \bibinfo{person}{Xite Wang}.} \bibinfo{year}{2023}\natexlab{}.
\newblock \showarticletitle{Exploring the Design Space of Unsupervised Blocking with Pre-trained Language Models in Entity Resolution}. In \bibinfo{booktitle}{\emph{{ADMA}}}, Vol.~\bibinfo{volume}{14176}. \bibinfo{pages}{228--244}.
\newblock


\bibitem[Thirumuruganathan et~al\mbox{.}(2021)]%
        {DBLP:journals/pvldb/Thirumuruganathan21}
\bibfield{author}{\bibinfo{person}{Saravanan Thirumuruganathan}, \bibinfo{person}{Han Li}, \bibinfo{person}{Nan Tang}, \bibinfo{person}{Mourad Ouzzani}, \bibinfo{person}{Yash Govind}, \bibinfo{person}{Derek Paulsen}, \bibinfo{person}{Glenn Fung}, {and} \bibinfo{person}{AnHai Doan}.} \bibinfo{year}{2021}\natexlab{}.
\newblock \showarticletitle{Deep Learning for Blocking in Entity Matching: {A} Design Space Exploration}.
\newblock \bibinfo{journal}{\emph{PVLDB}} \bibinfo{volume}{14}, \bibinfo{number}{11} (\bibinfo{year}{2021}), \bibinfo{pages}{2459--2472}.
\newblock


\bibitem[Vesdapunt et~al\mbox{.}(2014)]%
        {DBLP:journals/pvldb/VesdapuntBD14}
\bibfield{author}{\bibinfo{person}{Norases Vesdapunt}, \bibinfo{person}{Kedar Bellare}, {and} \bibinfo{person}{Nilesh~N. Dalvi}.} \bibinfo{year}{2014}\natexlab{}.
\newblock \showarticletitle{Crowdsourcing Algorithms for Entity Resolution}.
\newblock \bibinfo{journal}{\emph{PVLDB}} \bibinfo{volume}{7}, \bibinfo{number}{12} (\bibinfo{year}{2014}), \bibinfo{pages}{1071--1082}.
\newblock


\bibitem[Wang and Zhang(2024)]%
        {wang2024pre}
\bibfield{author}{\bibinfo{person}{Runhui Wang} {and} \bibinfo{person}{Yongfeng Zhang}.} \bibinfo{year}{2024}\natexlab{}.
\newblock \showarticletitle{Pre-trained Language Models for Entity Blocking: A Reproducibility Study}. In \bibinfo{booktitle}{\emph{NAACL}}. \bibinfo{pages}{8712--8722}.
\newblock


\bibitem[Wang et~al\mbox{.}(2020)]%
        {DBLP:conf/nips/WangW0B0020}
\bibfield{author}{\bibinfo{person}{Wenhui Wang}, \bibinfo{person}{Furu Wei}, \bibinfo{person}{Li Dong}, \bibinfo{person}{Hangbo Bao}, \bibinfo{person}{Nan Yang}, {and} \bibinfo{person}{Ming Zhou}.} \bibinfo{year}{2020}\natexlab{}.
\newblock \showarticletitle{MiniLM: Deep Self-Attention Distillation for Task-Agnostic Compression of Pre-Trained Transformers}. In \bibinfo{booktitle}{\emph{NIPS}}.
\newblock


\bibitem[Whang et~al\mbox{.}(2013)]%
        {DBLP:journals/tkde/WhangMG13}
\bibfield{author}{\bibinfo{person}{Steven~Euijong Whang}, \bibinfo{person}{David Marmaros}, {and} \bibinfo{person}{Hector Garcia{-}Molina}.} \bibinfo{year}{2013}\natexlab{}.
\newblock \showarticletitle{Pay-As-You-Go Entity Resolution}.
\newblock \bibinfo{journal}{\emph{TKDE}} \bibinfo{volume}{25}, \bibinfo{number}{5} (\bibinfo{year}{2013}), \bibinfo{pages}{1111--1124}.
\newblock


\bibitem[Yang et~al\mbox{.}(2019)]%
        {DBLP:conf/nips/YangDYCSL19}
\bibfield{author}{\bibinfo{person}{Zhilin Yang}, \bibinfo{person}{Zihang Dai}, \bibinfo{person}{Yiming Yang}, \bibinfo{person}{Jaime~G. Carbonell}, \bibinfo{person}{Ruslan Salakhutdinov}, {and} \bibinfo{person}{Quoc~V. Le}.} \bibinfo{year}{2019}\natexlab{}.
\newblock \showarticletitle{XLNet: Generalized Autoregressive Pretraining for Language Understanding}. In \bibinfo{booktitle}{\emph{NeurIPS}}. \bibinfo{pages}{5754--5764}.
\newblock


\bibitem[Zeakis et~al\mbox{.}(2023)]%
        {DBLP:journals/pvldb/ZeakisPSK23}
\bibfield{author}{\bibinfo{person}{Alexandros Zeakis}, \bibinfo{person}{George Papadakis}, \bibinfo{person}{Dimitrios Skoutas}, {and} \bibinfo{person}{Manolis Koubarakis}.} \bibinfo{year}{2023}\natexlab{}.
\newblock \showarticletitle{Pre-trained Embeddings for Entity Resolution: An Experimental Analysis}.
\newblock \bibinfo{journal}{\emph{PVLDB}} \bibinfo{volume}{16}, \bibinfo{number}{9} (\bibinfo{year}{2023}), \bibinfo{pages}{2225--2238}.
\newblock


\bibitem[Zecchini et~al\mbox{.}(2023)]%
        {DBLP:journals/pvldb/ZecchiniSBN23}
\bibfield{author}{\bibinfo{person}{Luca Zecchini}, \bibinfo{person}{Giovanni Simonini}, \bibinfo{person}{Sonia Bergamaschi}, {and} \bibinfo{person}{Felix Naumann}.} \bibinfo{year}{2023}\natexlab{}.
\newblock \showarticletitle{BrewER: Entity Resolution On-Demand}.
\newblock \bibinfo{journal}{\emph{PVLDB}} \bibinfo{volume}{16}, \bibinfo{number}{12} (\bibinfo{year}{2023}), \bibinfo{pages}{4026--4029}.
\newblock


\end{thebibliography}
\newpage

\section*{Appendix}
\label{sec:appendix}

\begin{figure*}[t]
  \centering
  \includegraphics[width=0.9\linewidth]{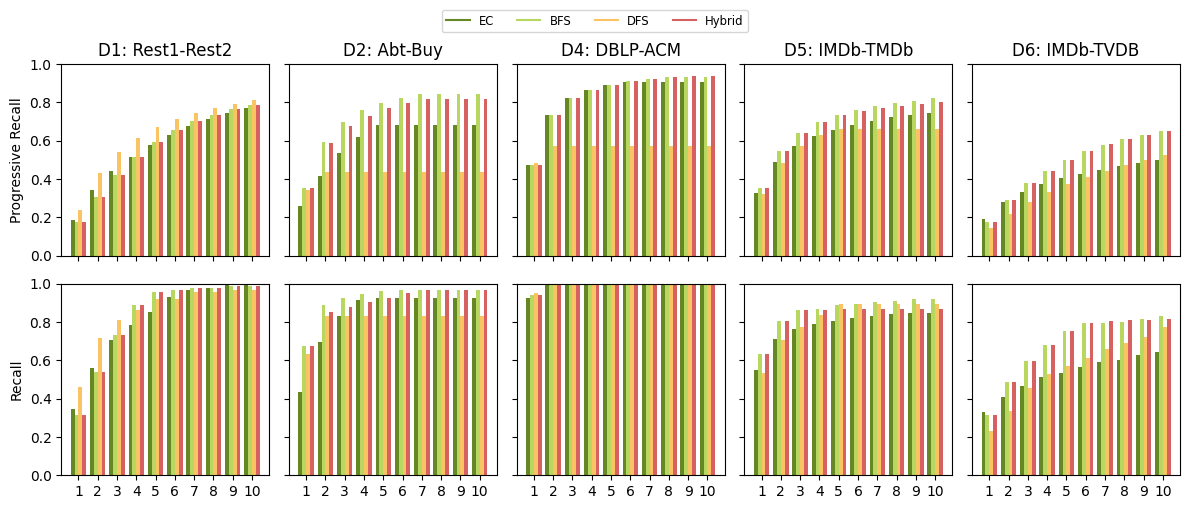}
  \vspace{-10pt}
  \caption{Progressive recall and recall of the best NN workflows in Table \ref{tb:nnConf}(a) across all budgets over the remaining Record Linkage datasets.}
  \label{fig:nnAUC2}
  \includegraphics[width=0.9\linewidth]{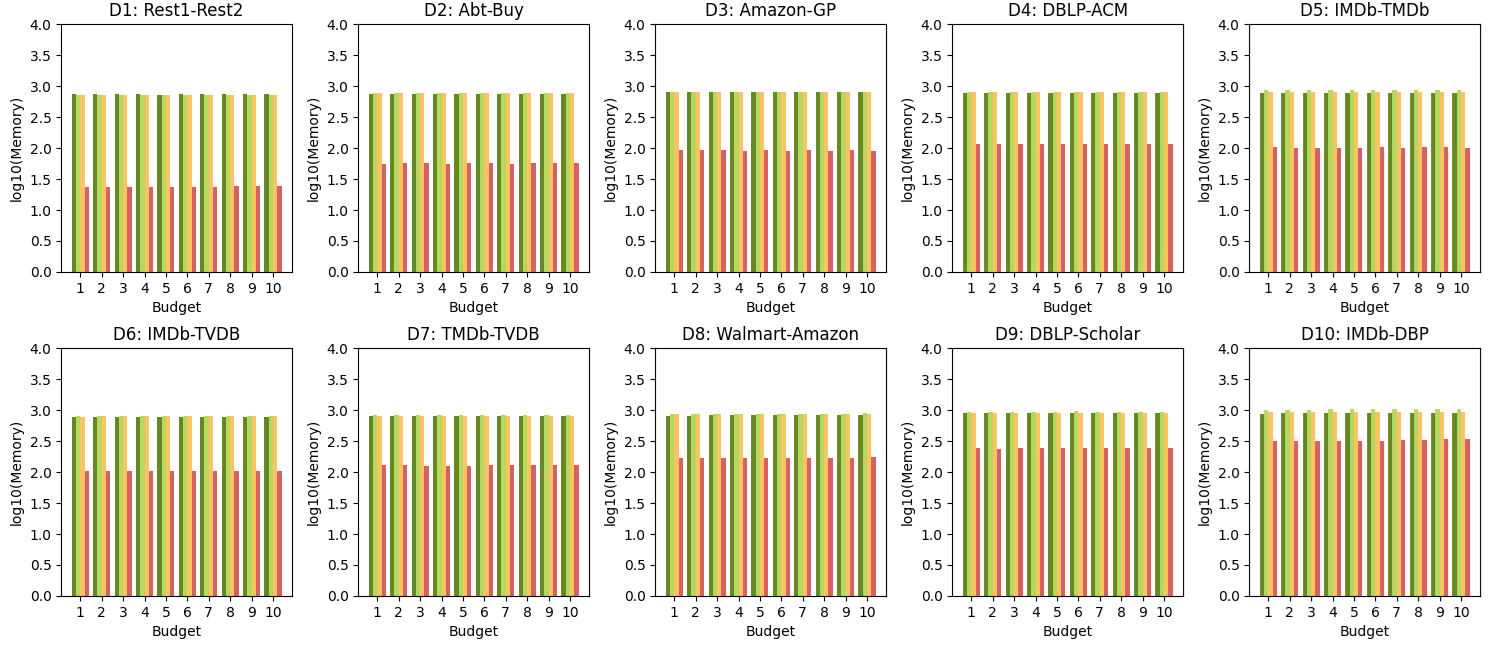}
  \vspace{-10pt}
  \caption{Memory footprint of the best NN workflows in Table \ref{tb:nnConf}(a) across all budgets over all Record Linkage datasets in Table \ref{tb:ccerDatasets}. The vertical axis is logarithmic and corresponds to MBs.}
  \label{fig:nnMemory}
  \includegraphics[width=0.9\linewidth]{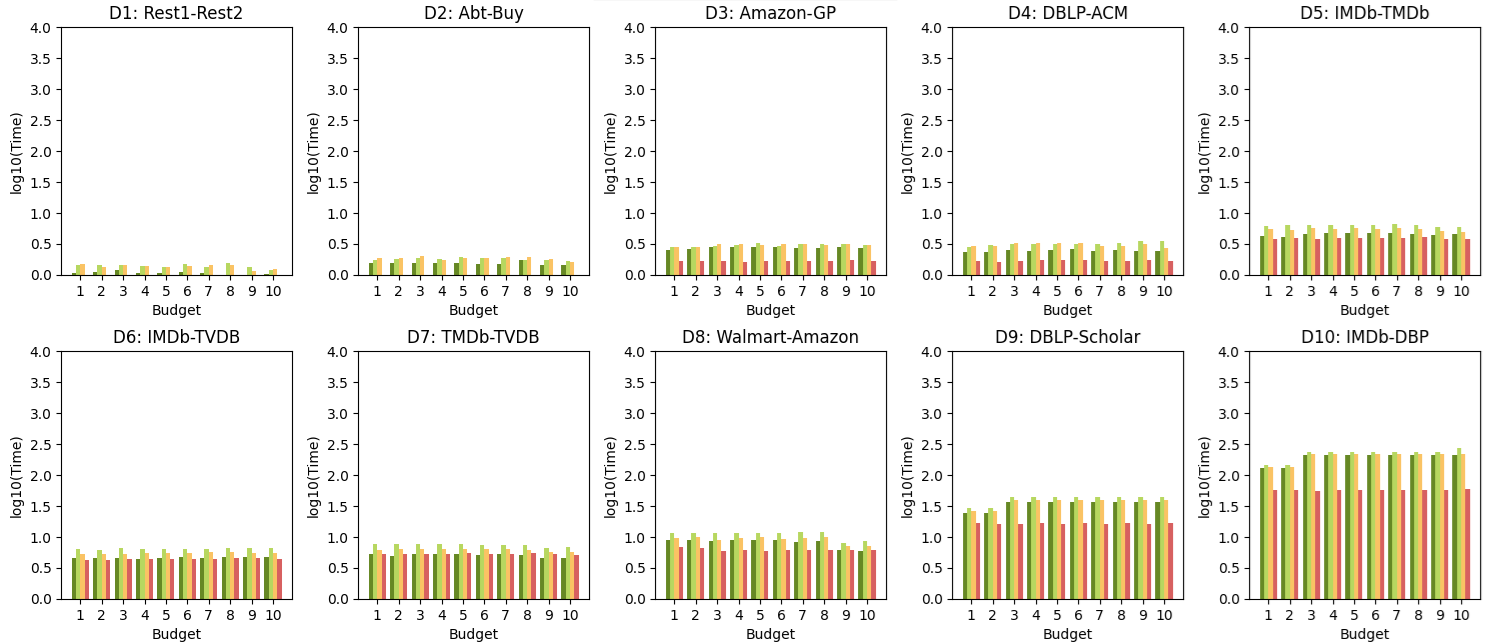}
  \vspace{-10pt}
  \caption{Run-times of the best NN workflows in Table \ref{tb:nnConf}(a) across all budgets over all Record Linkage datasets in Table \ref{tb:ccerDatasets}. The vertical axis is logarithmic and corresponds to seconds.}
  \label{fig:nn_time}
\end{figure*}

\begin{figure*}[t]
  \centering
  \includegraphics[width=0.9\linewidth]{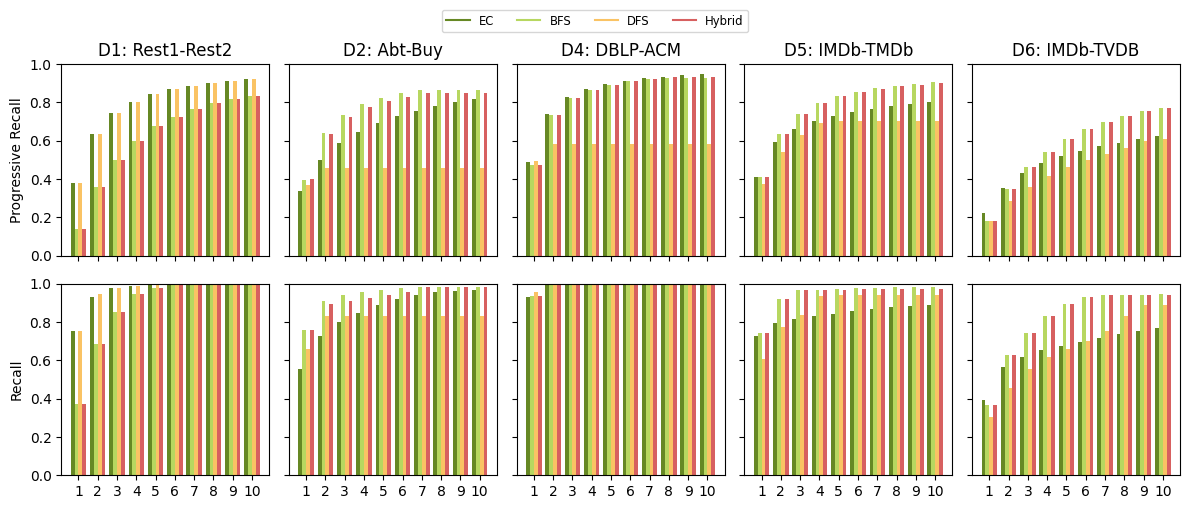}
  \vspace{-10pt}
  \caption{Progressive recall and recall of the best join workflows in Table \ref{tb:nnConf}(a) across all budgets over the remaining Record Linkage datasets.}
  \label{fig:joinAUC2}
  \includegraphics[width=0.9\linewidth]{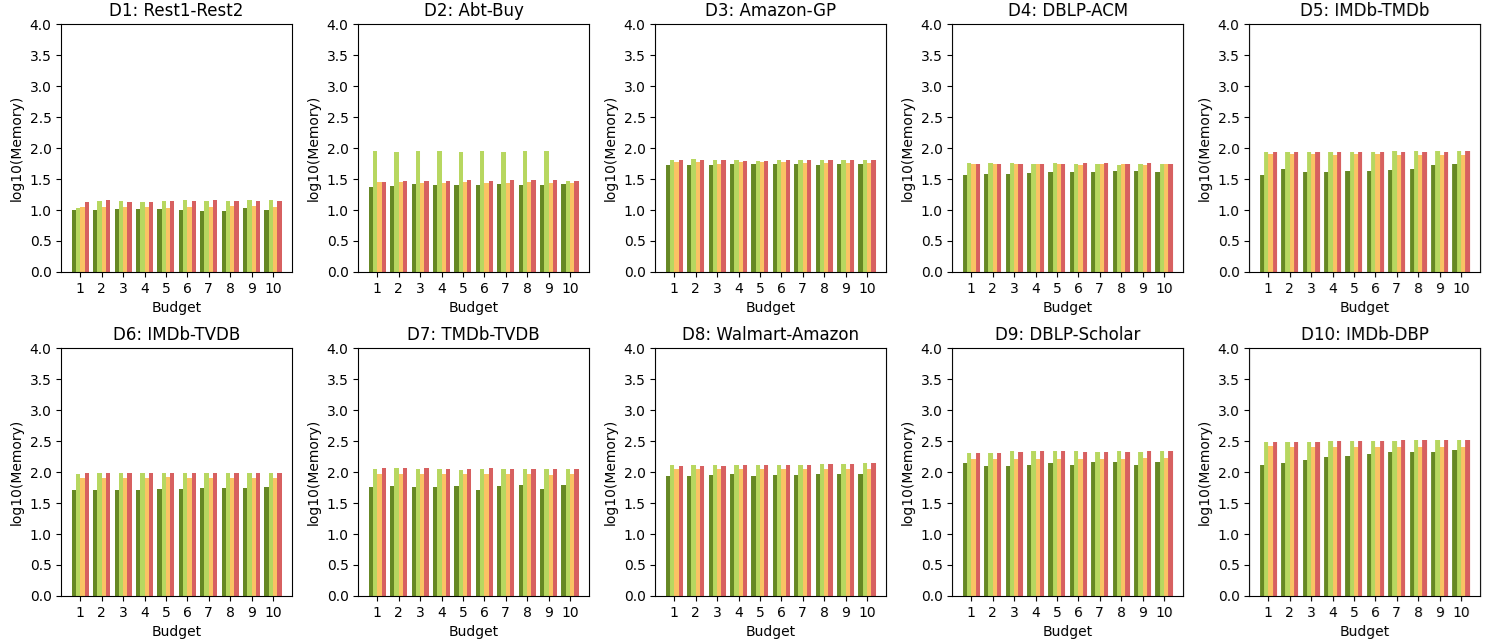}
  \vspace{-10pt}
  \caption{Memory footprint of the best join workflows in Table \ref{tb:nnConf}(a) across all budgets over all Record Linkage datasets in Table \ref{tb:ccerDatasets}. The vertical axis is logarithmic and corresponds to MBs.}
  \label{fig:joinMemory}
  \includegraphics[width=0.9\linewidth]{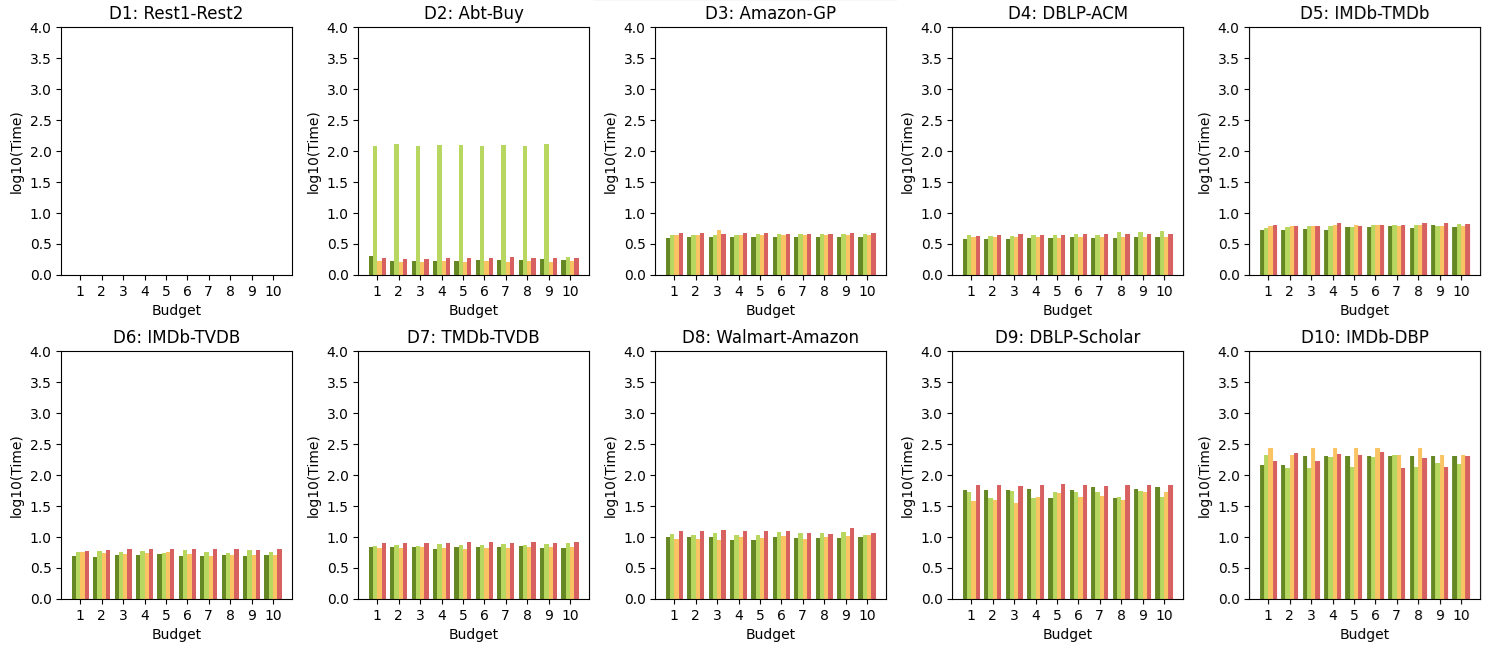}
  \vspace{-10pt}
  \caption{Run-times of the best join workflows in Table \ref{tb:nnConf}(a) across all budgets over all Record Linkage datasets in Table \ref{tb:ccerDatasets}. The vertical axis is logarithmic and corresponds to seconds.}
  \label{fig:join_time}
\end{figure*}

\begin{figure*}[t]
  \centering
  \includegraphics[width=0.9\linewidth]{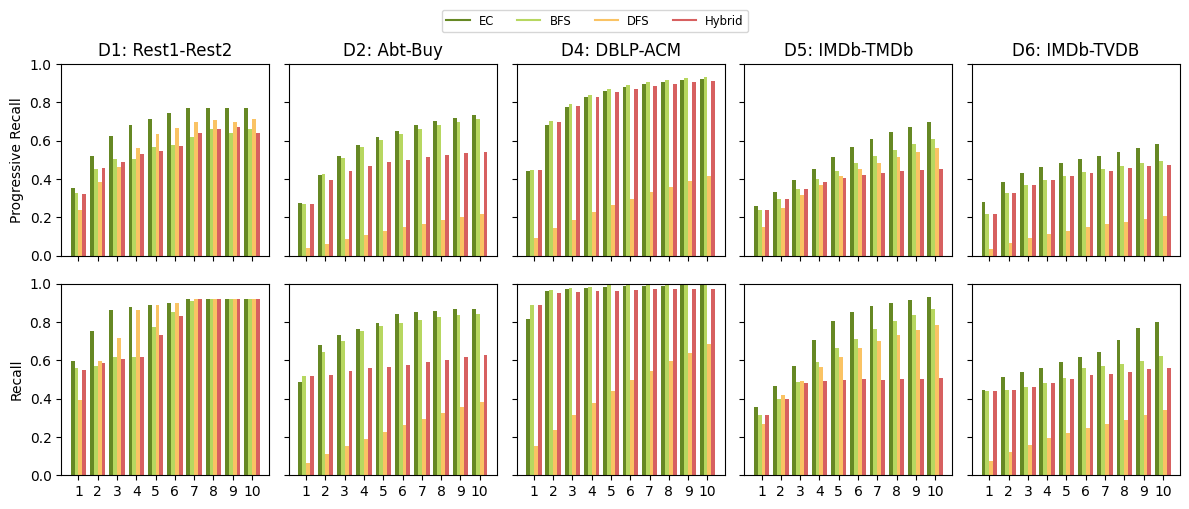}
  \vspace{-10pt}
  \caption{Progressive recall and recall of the best blocking workflows in Table \ref{tb:nnConf}(a) across all budgets over the remaining Record Linkage datasets.}
  \label{fig:blockAUC2}
  \includegraphics[width=0.9\linewidth]{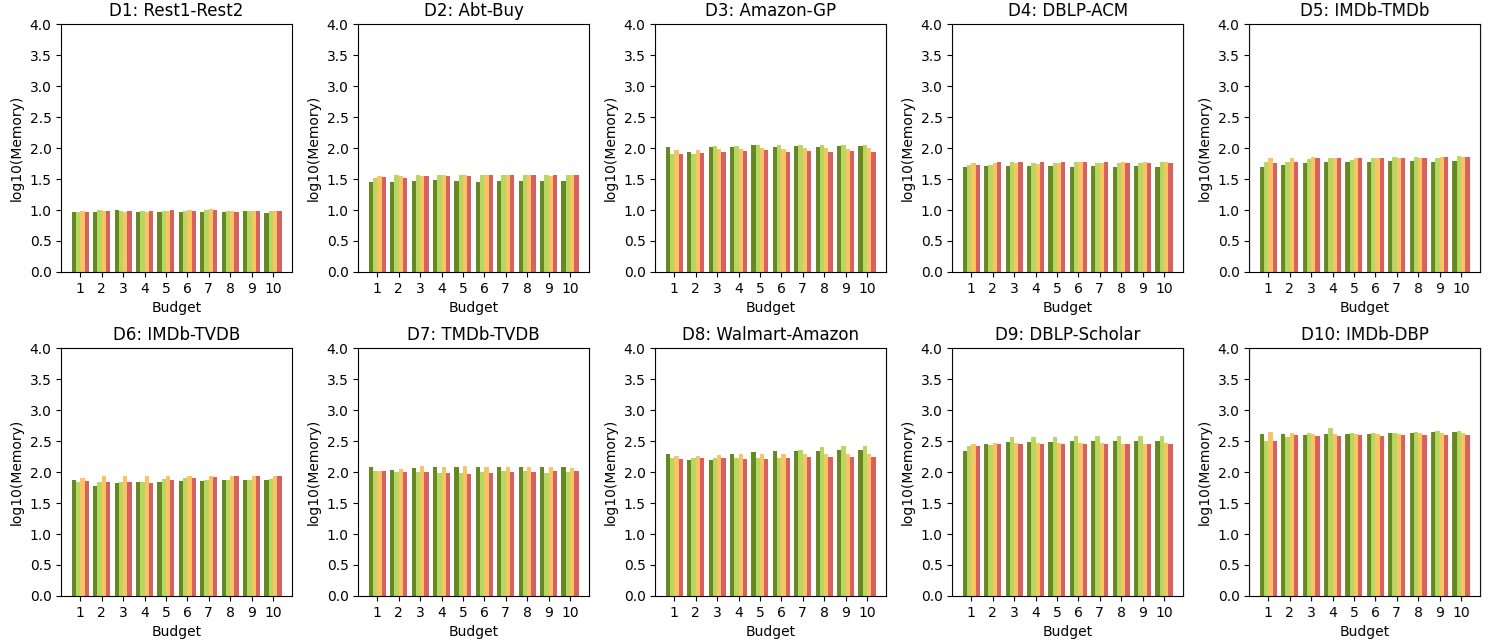}
  \vspace{-10pt}
  \caption{Memory footprint of the best blocking workflows in Table \ref{tb:nnConf}(a) across all budgets over all Record Linkage datasets in Table \ref{tb:ccerDatasets}. The vertical axis is logarithmic and corresponds to MBs.}
  \label{fig:pesmMemory}
  \includegraphics[width=0.9\linewidth]{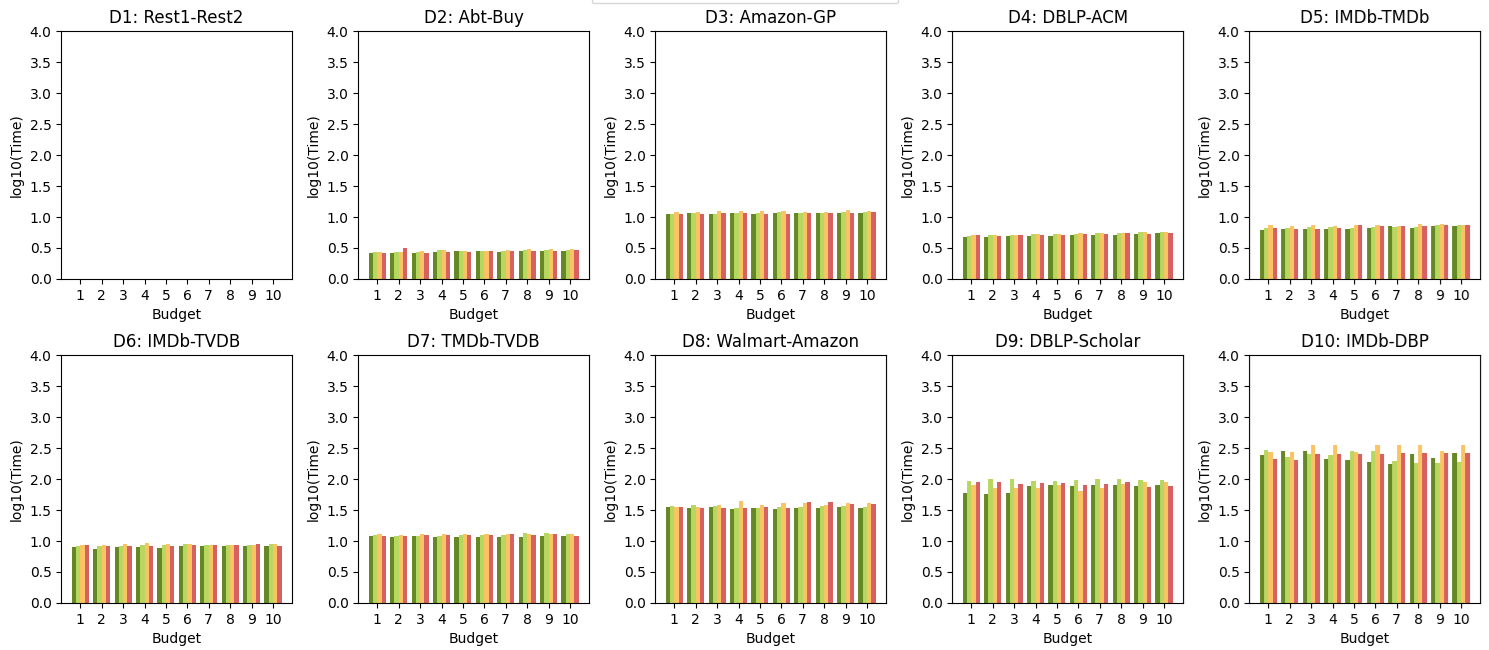}
  \vspace{-10pt}
  \caption{Run-times of the best blocking workflows in Table \ref{tb:nnConf}(a) across all budgets over all Record Linkage datasets in Table \ref{tb:ccerDatasets}. The vertical axis is logarithmic and corresponds to seconds.}
  \label{fig:pesm_time}
\end{figure*}

\begin{figure*}[t]
  \centering
  \includegraphics[width=0.9\linewidth]{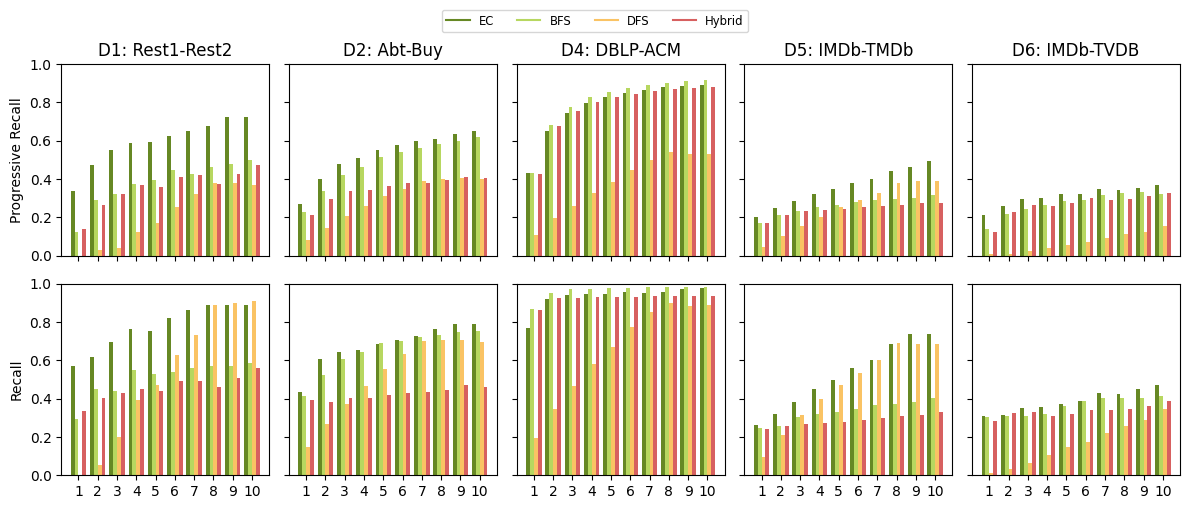}
  \vspace{-10pt}
  \caption{Progressive recall and recall of the best sorting-based workflows in Table \ref{tb:nnConf}(a) across all budgets over the remaining Record Linkage datasets.}
  \label{fig:snAUC2}
  \includegraphics[width=0.9\linewidth]{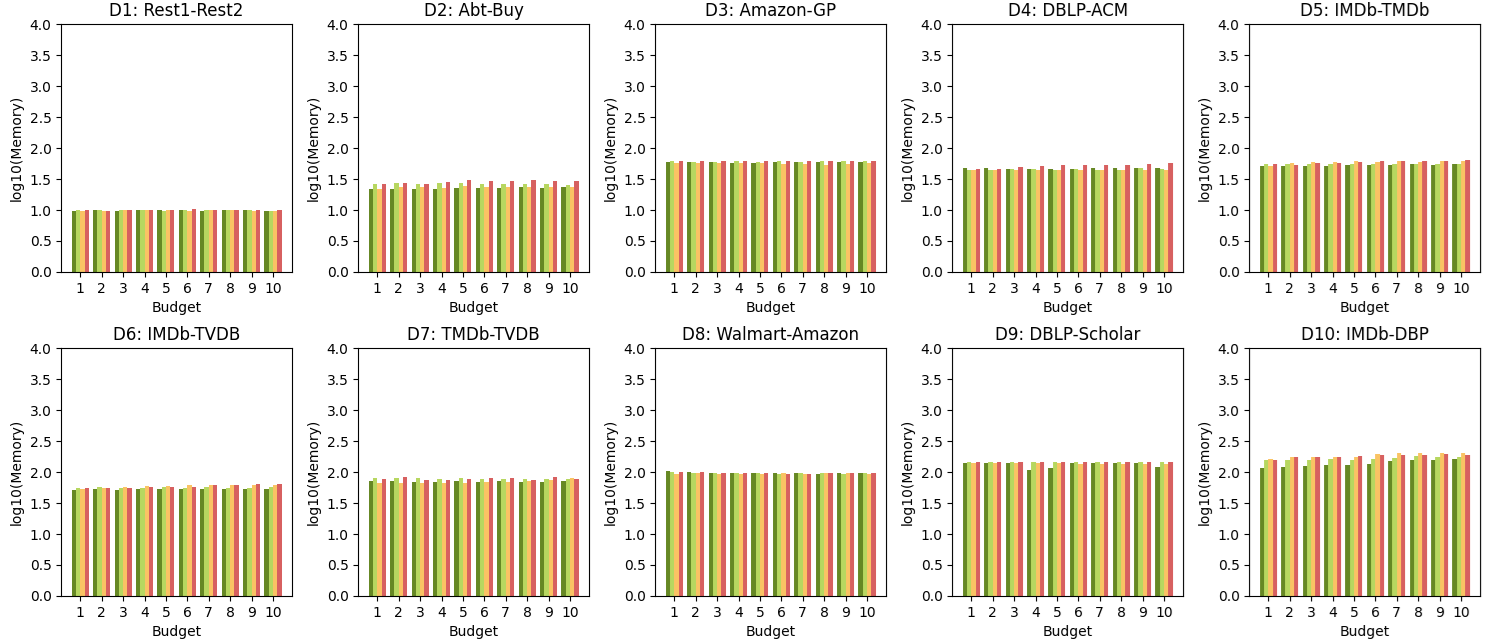}
  \vspace{-10pt}
  \caption{Memory footprint of the best sorting-based workflows in Table \ref{tb:nnConf}(a) across all budgets over all Record Linkage datasets in Table \ref{tb:ccerDatasets}. The vertical axis is logarithmic and corresponds to MBs.}
  \label{fig:snMemory}
  \includegraphics[width=0.9\linewidth]{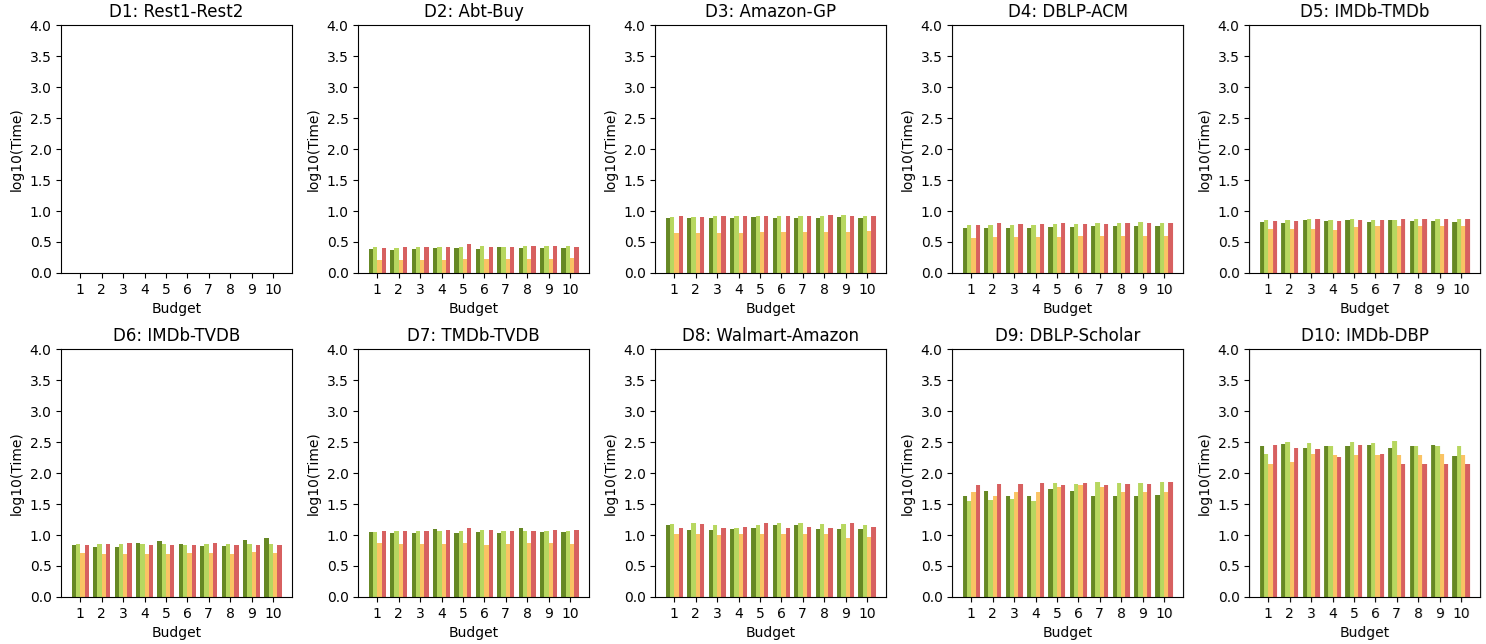}
  \vspace{-10pt}
  \caption{Run-times of the best sorting-based workflows in Table \ref{tb:nnConf}(a) across all budgets over all Record Linkage datasets in Table \ref{tb:ccerDatasets}. The vertical axis is logarithmic and corresponds to seconds.}
  \label{fig:sn_time}
\end{figure*}

\textbf{Datasets Description.} Starting with the Record Linkage datasets, D$_1$ was introduced in OAEI 2010\footnote{\url{http://oaei.ontologymatching.org/2010}}, matching restaurant descriptions. D$_2$ matches products from two online retailers: Abt.com and Buy.com \cite{DBLP:journals/pvldb/KopckeTR10}. D$_3$ comes from the same domain, matching products from Amazon.com and the Google Base data API (GB) \cite{DBLP:journals/pvldb/KopckeTR10}. D$_4$ is a bibliographic dataset, matching entries from DBLP with that of ACM Library \cite{DBLP:journals/pvldb/KopckeTR10}. D$_5$, D$_6$ and D$_7$ entail descriptions of television shows from TheTVDB.com (TVDB) and of movies from IMDb and themoviedb.org (TMDb) \cite{DBLP:conf/esws/ObraczkaSR21}.  D$_8$ is another product matching dataset, with data from Walmart and Amazon \cite{DBLP:conf/sigmod/MudgalLRDPKDAR18}. D$_9$ is bibliographic dataset with publications descriptions from DBLP and Google Scholar (GS) \cite{DBLP:journals/pvldb/KopckeTR10}.  D$_{10}$ matches movie descriptions from IMDb and DBpedia \cite{DBLP:journals/is/PapadakisMGSTGB20} -- the former has no overlap with the IMDb movies in D$_5$ and D$_6$. 

The Deduplication datasets are distinguished into real and synthetic ones. The former include Cora, a popular bibliographic dataset with machine learning publications \cite{DBLP:journals/pvldb/VesdapuntBD14}, CDdb, a set of audio CD descriptions from f freeDB\footnote{\url{http://www.freedb.org}} \cite{DBLP:journals/tkde/PapenbrockHN15} and Product, which is the original, dirty version of the product descriptions from Abt.com and Buy.com used in $D_2$ \cite{DBLP:journals/pvldb/KopckeTR10}. The five synthetic datasets have been generated with Febrl\footnote{\url{https://users.cecs.anu.edu.au/~Peter.Christen/Febrl/febrl-0.3/febrldoc-0.3/manual.html}} using the following standard approach \cite{DBLP:journals/pvldb/0001SGP16}: Initially, duplicate-free entities were extracted from frequency tables for real names (given and surname) and addresses (street number, name, postcode, suburb, and state names). Subsequently, duplicates were randomly created based on real error characteristics and modifications (e.g., inserting, deleting or substituting characters or words). The resulting structured, Deduplication datasets contain 40\% duplicate entities with up to 9 matches per entity, no more than 3 modifications per attribute value, and up to 10 modifications per entity.

\end{document}